\documentclass[prx,twocolumn,superscriptaddress,longbibliography]{revtex4-1}

\usepackage{graphicx,epsfig}
\usepackage{color}
\usepackage[usenames,dvipsnames]{xcolor}
\usepackage{amsmath}
\usepackage{bbm,amssymb, amsthm,array,hhline,tabularx,makecell}
\usepackage{dsfont} 
\usepackage{stmaryrd}
\usepackage{numprint}
\newcolumntype{Y}{>{\centering\arraybackslash}X}

\def\bb{\begin{eqnarray}}
\def\ee{\end{eqnarray}}
\newcommand{\ket}[1]{| #1 \rangle}
\newcommand{\bra}[1]{\langle #1 |}
\newcommand{\moy}[1]{\left\langle #1 \right\rangle}

\begin{document}

\title{Thermodynamics of Optical Bloch Equations}

\author{Cyril Elouard}
\email{cyril.elouard@gmail.com}
\affiliation{Department of Physics and Astronomy, University of Rochester, Rochester, NY 14627, USA}

\author{David Herrera-Mart\'i}
\affiliation{Atos Centre for Excellence in Performance Programming, 1 rue de Provence, 38130 Echirolles, France}

\author{Massimiliano Esposito}
\email{massimiliano.esposito@uni.lu}
\affiliation{Complex Systems and Statistical Mechanics, Department of Physics and Materials Science, University of Luxembourg, L-1511 Luxembourg, Luxembourg}

\author{Alexia Auff\`eves}
\email{alexia.auffeves@neel.cnrs.fr}
\affiliation{CNRS and Universit\'e Grenoble Alpes, Institut N\'eel, F-38042 Grenoble, France}

\date{\today}

\begin{abstract}
Optical Bloch Equations (OBE) describe the coherent exchange of energy between a quantum bit (qubit) and a quasi-resonant driving field in the presence of a thermal bath. Despite it being an ubiquitous process in quantum technologies, a sound thermodynamic analysis is still missing. We hereby provide such an analysis, by deriving the relevant framework from first principles. We start from a complete microscopic description of the qubit-bath system where definitions of heat, work and entropy production are unambiguous. We trace out the bath and coarse-grain the resulting expressions in time, using a methodology similar to the derivation of the dynamical master equation, to derive closed expressions for the First and Second Law in terms of system properties. Long coarse graining times yield the Floquet Master Equation and its already known thermodynamic description. Short coarse-graining times yield instead to the OBE and a novel thermodynamic framework which explicitly depend on quantum coherences in the qubit's energy basis which produce quantum signatures in the heat and entropy production flows. This allow us to characterize a genuinely quantum non-equilibrium situation, where the coherences created by the driving field are continuously erased by the bath. Our findings can be readily extended to larger open quantum systems. They carry the seeds for future thermodynamic analyses of quantum gates and the design of quantum engines in the strong coherent driving regime.
\end{abstract}

\maketitle

\section{Introduction}

Quantum thermodynamics is an emerging field at the interface between quantum optics, quantum information and stochastic thermodynamics, that aims to extend the laws of thermodynamics in the quantum regime \cite{Vinjanampathy16,QTSpringer}. Its motivations can be fundamental -- are there quantum sources of irreversibility \cite{Santos19,Mohammady20}? can we design quantum engines with coherence enhanced performances \cite{Herrera-Marti15,Uzdin15,Uzdin16,Klatzow19,Holmes20,Watanabe20}?-- or practical -- what is the energetic cost of coherently manipulating a quantum system, or stabilizing a quantum state against noise \cite{Elouard17Role}? 

A common thread in those new and exciting questions is the identification of ``quantum signatures", i.e. thermodynamic clues of the presence of quantum coherences in the state of the system of interest. To single out those signatures, a highly desirable regime is the so-called coherent regime of driving: A strong, time-dependent field drives some quasi-resonant transition in a quantum system. Simultaneously, this system interacts with a bath that perturbs the coherent dynamics. In the canonical case of a quantum bit or qubit, this scenario has for long been successfully modeled by the textbook Optical Bloch Equations (OBE). The ``coherent regime of driving" defined above is manifested by the celebrated Rabi oscillation \cite{CCT}, where energy is coherently and reversibly exchanged between the field and the qubit. This mechanism is crucial for many functionalities of quantum technologies, e.g. the implementation of quantum gates based on the resonant addressing of qubits' transitions, or the amplification of light by stimulated emission.

Despite its importance for quantum protocols, the coherent regime of the OBE is still lacking a consistent thermodynamic description. A key issue is that the dissipation term it involves is similar to the one experienced by the bare (undriven) qubit, whichever the driving strength may be. For this reason, thermodynamic analyses have been performed in the weak regime of driving where the qubit dynamics is dominated by dissipation \cite{Uzdin15,Uzdin16,Klatzow19}. When the drive is strong, they have relied on Floquet Master Equations (FME) \cite{Alicki13,Langemeyer14,Cuetara15,Donvil18}, which involve the dressed qubit energy spectrum but require a coarse-graining in time that largely overcomes the Rabi period, blurring out coherent energy exchanges. Other thermodynamic analyses of quantum processes have focused on ideal cases where the unitary and incoherent evolutions are temporally separated \cite{Kosloff02,Alecce15,Thomas14,Pekola19,Landi20}.

In this article, we provide a thermodynamic analysis of the OBE in the coherent regime of driving. Our findings are based on a new strategy where the desired framework is derived from first principles. Building on a microscopic model of the bath, we first express heat, work and entropy production flows at the level of the closed qubit-bath system, where the definitions are unambiguous \cite{Esposito10Entropy}. We coarse-grain the obtained expressions using a methodology similar to the derivation of the dynamical master equation, to obtain consistent expressions of the First and Second Law at the qubit's level. This derivation is drastically different from former approaches where the thermodynamic definitions are postulated at the level of the already derived master equation \cite{Geva95,Alicki13,Hofer17}.

We have verified that the coarse graining time giving rise to the FME allows to recover the thermodynamic framework derived in  \cite{Alicki13,Langemeyer14,Cuetara15,Donvil18}. Conversely, using the coarse graining time that leads to the OBE brings the desired thermodynamic framework, whose variables explicitly depend on the quantum coherences in the qubit's energy basis. We finally exploit these quantum signatures in the heat and entropy production flows to characterize a non-equilibrium steady-state with no classical equivalent, where the coherences built by the drive in the bare qubit energy eigenbasis are continuously removed by the thermal bath. 


Our results go beyond the paradigmatic case of the OBE and can be generalized to a large class of driven quantum open systems, e.g. involving more than two levels as it is the case for weakly interacting multipartite open systems. 
They open the way towards thermodynamic analyzes of quantum gates and the design of heat engines in deep coherent regimes, beyond weak or slow driving \cite{Uzdin15,Klatzow19}.

Section \ref{s:ME} presents a unified derivation of the master equation ruling the qubit's dynamics. Building on the microscopic model of the bath, we show which choice of a coarse-graining time determines the emergence of the FME or the OBE, clarifying their respective regimes of validity. The former (resp. the latter) involves a coarse-graining time that is longer (resp. shorter) than the Rabi period. By the way, we show that despite their local nature, the OBE are fully consistent with the Second Law, making them suitable to define thermodynamic quantities.  We then switch to the thermodynamic analysis. From the description of the joint system-bath evolution, we define and compute in Section \ref{s:Thermo} the heat flow as the energy provided by the bath to the system. We then derive the First and Second Laws of thermodynamics for the qubit consistent with the timescales captured by the OBE and the FME, respectively. The study of the quantum non-equilibrium situation and its thermodynamic signatures is presented in Section \ref{s:Signatures}.


\section{Dynamical analysis}
\label{s:ME}

In this Section we derive two master equations ruling the dynamics of a driven qubit coupled to a thermal bath, namely the OBE and the FME. While these equations are well-known, the ensuing novel derivation is essential to precisely identify their regimes of validity and to properly formulate their respective thermodynamic behavior, as will be done in the next Section.

\subsection{Microscopic derivation}\label{s:Redfield}

\subsubsection{System and model}

\begin{figure*}
\begin{center}
\includegraphics[width=0.80\textwidth]{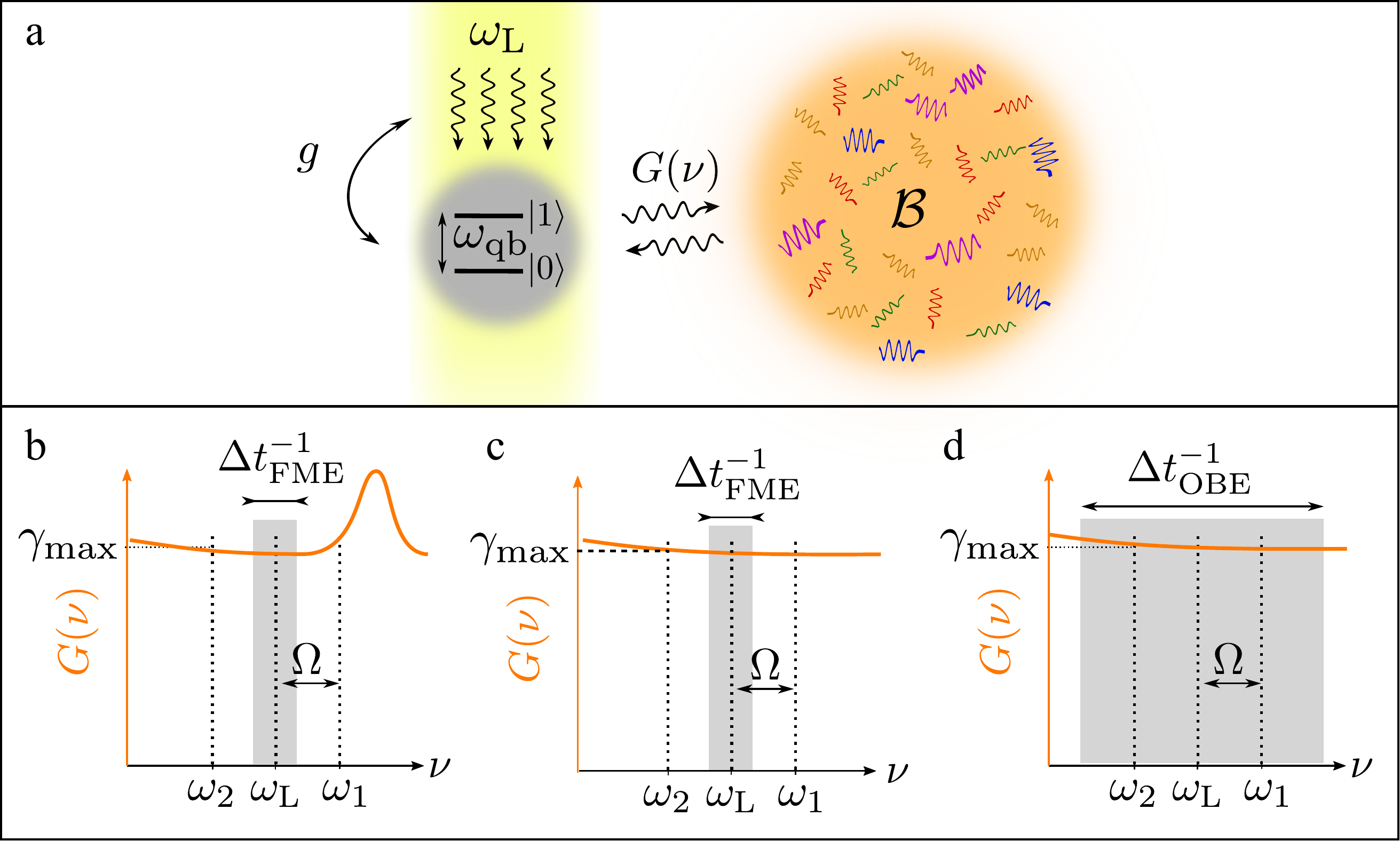}
\end{center}
\caption{\textbf{a}: 
{\bf Resonance fluorescence.} A qubit of frequency $\omega_\text{qb}$ is driven by a quasi-resonant monochromatic field of frequency $\omega_\text{L}$ with a coupling strength $g$. The qubit also interacts with a thermal bath $\cal B$ characterized by its temperature $T$ and spectral function $G(\nu)$ (See text). 
\textbf{b},\textbf{c},\textbf{d}: \textbf{Two possible choices of coarse-graining}. Dotted vertical lines: Fourier frequencies of the driven qubit operators $\sigma_\pm^I(t)$ (See text), i.e. $\omega_L$, $\omega_1 = \omega_\text{L}+\Omega$ and $\omega_2 = \omega_\text{L}-\Omega$, with $\Omega$ the Rabi frequency. Orange solid line: Spectral density $G(\nu)$, with $\gamma_\text{max}$ its maximum value over the set of Fourier frequencies. The bath correlation time $\tau_c$ scales like the inverse of the width of $G(\nu)$. \textbf{b}: $\tau_c  \gtrsim \Omega^{-1}$. The coarse-graining time must verify $\Delta t_\text{FME} \gg \Omega^{-1}$, leading to the emergence of FME.  \textbf{c,d}: $\tau_c  \lesssim\Omega^{-1}$. The choice of a long (resp. a short) coarse-graining time $\Delta t_\text{FME} \gg \Omega^{-1}$ (resp. $\Delta t_\text{OBE} \ll \Omega^{-1}$) in \textbf{c} (resp. \textbf{d}) leads to the emergence of the FME (resp. the OBE), see text.} \label{f:Scenery}
\end{figure*}

We consider a driven qubit of frequency denoted $\omega_\text{qb}$  (See Fig.\ref{f:Scenery}\textbf{a}). The qubit dynamics is governed by the Hamiltonian $H(t) = H_\text{qb} + V(t)$, where ${H_\text{qb} = \hbar \omega_\text{qb}\sigma_z/2}$ is the free Hamiltonian of the qubit. Conversely, ${V(t) = \frac{\hbar g}{2}\cos(\omega_\text{L}t)\sigma_x}$ describes the driving by a quasi-resonant classical field of frequency ${\omega_\text{L} = \omega_\text{qb} - \delta}$ where $\vert\delta\vert\ll\omega_\text{qb}$ stands for the qubit-field detuning. We introduced the Pauli matrices $\sigma_z = \ket{1}\bra{1}-\ket{0}\bra{0}$ and $\sigma_x = \ket{1}\bra{0} + \ket{0}\bra{1}$, where $\ket{0}$ and $\ket{1}$ are the ground and excited state of the qubit respectively. In the rest of the paper, we shall use a terminology imported from atomic physics where the mechanism of resonance fluorescence was first studied. In particular, we shall refer to the qubit's dipole oscillating in phase (resp. in quadrature of phase) with the drive as the observable associated with $\sigma_x$ (resp. $\sigma_y$). The qubit -- field coupling strength verifies ${g \ll \omega_\text{qb}, \omega_\text{L}}$. This gives ground for working within the Rotating Wave Approximation \cite{CCT} such that ${V(t) \simeq \hbar g\left(e^{i\omega_\text{L}t}\sigma_- + e^{-i\omega_\text{L}t} \sigma_+\right)}$.

The qubit is coupled to a photonic bath ${\cal B}$ modeled by a collection of bosonic modes of respective frequencies $\omega_k$, lowering operators $a_k$ and free Hamiltonian $H_{\cal B} = \sum_k \hbar\omega_k \left(a^\dagger_k a_k + 1/2\right)$. The bath is assumed to be in a thermal equilibrium state $\rho_{\cal B}^\text{eq} = e^{-H_{\cal B}/k_BT}/Z_{\cal B}$ where $T$ stands for the bath temperature and $Z_{\cal B}$ the partition function. Finally, the qubit-bath coupling is described by the Hamiltonian $H_{\cal SB} = R \sigma_x$, where $R = \sum_k \hbar g_k \left(a_k^\dagger +a_k\right)$ is a bath operator. The parameters $g_k$ are taken real without loss of generality. The density operator $\rho_{\cal SB}$ of the joint qubit-bath system obeys the exact Liouville-Von Neumann equation:
\bb
\dot{\rho}_{\cal SB}(t) = -\dfrac{i}{\hbar}\left[H(t) + {H}_{\cal SB}(t) + H_{\cal B},\rho_{\cal SB}(t)\right]\label{eq:VNE}.
\ee
Both the OBE and the FME rely on the assumption that the coupling between the bath and the driven qubit is weak enough such that the total density operator ${\rho}_{\cal SB}$ evolves slowly when written in the appropriate interaction picture. This new frame is reached via two successive transformations. The first transformation is generated by the unitary operator $U_\text{r} = e^{i\omega_\text{L}t\sigma_z/2}$ and allows us to describe the dynamics in the frame rotating at the drive frequency. Operators in this representation being denoted by a tilde, the qubit-bath density operator $\tilde\rho_{\cal SB}(t) = U_\text{r}\rho_{\cal SB}(t)U_\text{r}^\dagger$ verifies the evolution equation 
\bb
\dot{\tilde\rho}_{\cal SB}(t) = -\dfrac{i}{\hbar}\left[\tilde{H}_\text{eff} + \tilde{H}_{\cal SB}(t) + H_{\cal B},\tilde\rho_{\cal SB}(t)\right],\label{eq:VNEti}
\ee
$\tilde H_\text{eff} = U_\text{r}H(t)U_\text{r}^\dagger  -\tfrac{\hbar\omega_\text{L}}{2}\sigma_z= \frac{\hbar\delta}{2}\sigma_\text{z} + \frac{\hbar g}{2}\sigma_x$ is the effective Hamiltonian of the qubit in the rotating frame and $\tilde H_{\cal SB}(t) =  R\left(\sigma_-e^{-i\omega_\text{L}t} + \sigma_+e^{i\omega_\text{L}t} \right)$. The eigenstates of $\tilde H_\text{eff}$ are the standard dressed qubit states
\begin{subequations}
\bb
\ket{+} &=& \frac{\sqrt{\Omega+\delta}}{\sqrt{2\Omega}}\ket{1} + \frac{\sqrt{\Omega-\delta}}{\sqrt{2\Omega}}\ket{0}\label{d:plus}\\
\ket{-} &=& -\frac{\sqrt{\Omega-\delta}}{\sqrt{2\Omega}}\ket{1} + \frac{\sqrt{\Omega+\delta}}{\sqrt{2\Omega}}\ket{0}\label{d:minus}.
\ee
\end{subequations}
$\ket{\pm}$ are respectively associated with the eigenvalues $\pm \hbar \Omega/2$, where 
\bb
\Omega = \sqrt{g^2+\delta^2}\label{d:RabiFreq}
\ee
is the so-called Rabi frequency. In a second step, we introduce the interaction picture with respect to $\tilde H_\text{eff}+ H_{\cal B}$, that is defined by the unitary operator $U_\text{int} = e^{it(\tilde H_\text{eff}+ H_{\cal B})/\hbar}$. We use the superscript $I$ to denote the result of the two successive transformations. The evolution of $\rho_{\cal SB}^I(t) = U_\text{int} \tilde \rho_{\cal SB}(t) U_\text{int}^\dagger$ is governed by $H_{\cal SB}^I(t) = U_\text{int} \tilde{H}_{\cal SB}(t) U_\text{int}^\dagger$:
\bb
\dot{\rho}^I_{\cal SB}(t) = -\dfrac{i}{\hbar}\left[H_{\cal SB}^I(t),\rho_{\cal SB}^I(t)\right].\label{eq:VNEI}
\ee
Finally, it is convenient to rewrite $H^I_{\cal SB}(t)$ by identifying the Fourier components of $\sigma_l^I(t) = e^{il\omega_\text{L}t}\sum_{\omega =0,\pm\Omega} \sigma_l(\omega)e^{i\omega t}$, for $l=\pm$, in terms of the eigenstates of $\tilde H_\text{eff}$. We obtain:

\begin{subequations}
\bb
\sigma_\pm(0) &=& \dfrac{g}{2\Omega}\big(\ket{+}\bra{+} - \ket{-}\bra{-}\big)\label{d:sigpm0}\\
\sigma_\pm(-\Omega) &=& \mp \dfrac{\Omega\mp \delta}{2\Omega}\ket{-}\bra{+}\\
\sigma_\pm(\Omega) &=& \pm \dfrac{\Omega\pm \delta}{2\Omega}\ket{+}\bra{-},\label{d:sigpmp}
\ee
\end{subequations}
such that $H^I_{\cal SB}(t) = R^I(t)\sum_{l=\pm}\,\sum_{\omega={0,\pm\Omega}} \sigma_l(\omega)e^{il\omega_\text{L}t}e^{i\omega t}$.

\subsubsection{Properties of the bath} 

The typical features of the dissipation experienced by the driven qubit are related to the spectral density $G(\nu)$ of the bath, defined as the Fourier transform of the qubit-bath correlation function $\moy{R^I(\tau)R^I(0)} = \text{Tr}\{R^I(\tau)R^I(0)\rho_{\cal B}^\text{eq}\}$:
\bb
G(\nu) = \frac{1}{\hbar^2}\int_{-\infty}^\infty d\tau e^{i\nu\tau}\moy{R^I(\tau)R^I(0)}\label{d:Gnu}.
\ee
The bath correlation time $\tau_\text{c}$ corresponds to the decay time of the correlation function, and is thus set by the inverse width of the spectral density $G(\nu)$ \cite{Breuer}. 
Using the specific form of the operator $R$ and of the bath's equilibrium state $\rho^\text{eq}_{\cal B}$, we derive the explicit form of $G(\nu)$. Let us introduce $N(\nu) =[e^{\hbar\nu/k_BT}-1]^{-1}$ the thermal mean occupation of the mode at frequency $\nu$. Noting that  $\text{Tr}\{\rho_{\cal B}^\text{eq}a_j^\dagger a_{k}\}= \delta_{j,k}N(\omega_k)$, and that $\text{Tr}\{\rho_{\cal B}^\text{eq}a_j a_k\} = 0$, we get:
\bb
\frac{G(\nu)}{2} = \Theta(\nu)\Gamma(\nu)(N(\nu)+1)+ \Theta(-\nu)\Gamma(-\nu)N(-\nu)\nonumber\\
\ee
$\Theta(\nu)$ stands for the Heaviside function. We defined the zero-temperature bath spectral function
\bb
\Gamma(\nu) &\underset{}{=}& \sum_k g_k^2 \delta_\text{D}(\nu-\omega_k),
\ee
with $\delta_\text{D}(\nu)$ denoting the Dirac distribution. In the absence of driving, $\Gamma(\omega_\text{qb})$ is the qubit's spontaneous emission rate. $\Gamma(\nu)$ contains information about the physical structure of the bath, e.g. the presence of resonances being highlighted by Lorentzian peaks. Such resonances lead to bath memory effects and increased values of $\tau_c$. It is straightforward to check that $G(-\nu)  = e^{-\hbar\nu/k_\text{B}T}G(\nu)$, i.e. that $G(\nu)$ complies with the Kubo-Martin-Schwinger (KMS) condition \cite{Kubo57,Martin59}, as expected for a bath at thermal equilibrium. This plays a crucial role to ensure that the resulting master equation is thermodynamically consistent \cite{Breuer}.

\subsubsection{Coarse-graining procedure}

Our derivation of the driven-qubit master equation is based on a coarse-graining of the exact evolution equation of the driven qubit and the bath \cite{CCT}. The variation time of ${\rho}^I_{\cal SB}$ typically scales like $\gamma_\text{max}^{-1}$, where $\gamma_\text{max}$ is the maximum of $G(\nu)$ for $\nu$ taken within the Fourier frequencies of $H^I_{\cal SB}(t)$, namely $\{\pm\omega_\text{L},\pm\omega_\text{L}\pm\Omega\}$ (See Fig.\ref{f:Scenery}\textbf{b}- \textbf{d}). In the limit $\gamma_\text{max}^{-1} \gg \tau_\text{c}$, it is possible to coarse-grain in time Eq.~\eqref{eq:VNEI} over a time-scale 
\bb
\tau_\text{c}\ll \Delta t_\text{cg} \ll  \gamma_\text{max}^{-1}, \label{eq:Dtcg}
\ee
then trace over the bath degrees of freedom to derive a Markovian master equation for the driven qubit.
The procedure, presented generically in Ref.~\cite{CCT}, yields an equation similar to the Redfield equation \cite{Redfield65} for the density operator $\rho^I(t) = \text{Tr}_{\cal B}\{\rho^I_{\cal SB}(t)\}$ of the qubit:
\bb
\dot\rho^I(t) &=& -\frac{1}{\hbar^2\Delta t_\text{cg}}\!\displaystyle\int_{t}^{t+\Delta t_\text{cg}}\!\!\!\!\!\!\!\! dt'\!\!\int_{t}^{t'}\!\!\!\! dt''\nonumber\\
&&\quad\quad\text{Tr}_{\cal B}\!\Big\{\left[H^I_{\cal SB}(t'),\!\left[H^I_{\cal SB}(t''),\rho^I_{\cal SB}(t'')\right]\right]\!\!\Big\}.\quad
\label{eq:Redfield}
\ee
It was assumed without loss of generality that the expectation value of $H_{\cal SB}$ in the bath equilibrium state is zero \cite{Gaspard99,CCT}, such that no first order contribution in the coupling Hamiltonian remains. 
We further simplify Eq.~\eqref{eq:Redfield} by applying the Born-Markov approximation \cite{CCT,Breuer}: We replace $\rho^I_{\cal SB}(t'')$ by $\rho^I(t)\!\otimes\!\rho^\text{eq}_{\cal B}$ inside the double integration, assuming the bath to be in the equilibrium state $\rho^\text{eq}_{\cal B}$ and neglecting the (slow) evolution of the qubit density operator in the interaction picture over $\Delta t_\text{cg}$. Finally, we rewrite Eq.~\eqref{eq:Redfield} using the explicit expression of $H^I_{\cal SB}(t)$:
\bb
&&\dot\rho^I\!(t) =  \dfrac{1}{4\Delta t_\text{cg}}\int_t^{t+\Delta t_\text{cg}}\!\!\! dt' \sum_{ll'=\pm} \sum_{\omega,\omega'=0,\pm\Omega} \!\!\! e^{i(\omega- \omega')t'+i(l-l')\omega_\text{L}t'}\nonumber\\
&&\times G(-\omega - l\omega_\text{L})\!\left[\sigma_{l}(\omega)\rho^I(t) \sigma_{l'}^\dagger (\omega') -
 \sigma_{l'}^\dagger (\omega') \sigma_{l}(\omega)\rho^I\!(t)  \right]\nonumber\\
 &&+\, \text{H.c.}\label{eq:pre}
\ee
where $\text{H.c.}$ stands for Hermitic conjugate. To go to Eq.~\eqref{eq:pre}, the integral over $t''$ in Eq.~\eqref{eq:Redfield} was turned into an integral over $\tau = t'-t''$. Its upper limit can be approximated by $+\infty$ as the integrand contains the correlation function $\moy{R^I(0)R^I(\tau)}$ which decays for $\tau \gg \tau_\text{c}$ \cite{CCT,Breuer}. The imaginary part of $\int_0^\infty d\tau e^{i\nu\tau}\moy{R^I(\tau)R^I(0)}$ contributes to the Lamb and dynamic Stark shifts, which are small corrections to the free qubit frequency (See \cite{Farley81,Carmichael1} and Appendix A). They are omitted in the following.

Eq.~\eqref{eq:pre} is not ensured to be completely positive since it is not yet of Lindblad form \cite{Lindblad76}. The derivation therefore requires an additional step, called Secular Approximation \cite{CCT}, which consists in neglecting the terms oscillating with time $t'$ with a frequency larger than ${\Delta t_\text{cg}}^{-1}$. A positive master equation can be obtained with $\Delta t_\text{cg}\gg \omega_\text{L}^{-1},\omega_\text{qb}^{-1}$. Furthermore, the magnitude of the coarse-graining time determines the smallest timescale the master equation can describe, and its capacity to capture coherent energy exchanges. 
When $ \tau_c \geq \Omega^{-1}$, the coarse-graining time necessarily verifies $ \Delta t_\text{cg} \gg \Omega^{-1}$, such that the resulting master equation does not resolve the Rabi oscillations (see Fig.~\ref{f:Scenery}\textbf{b}). This corresponds to the FME. 
On the opposite, when $ \tau_c \leq \Omega^{-1}$, two choices can be made. Keeping $\Delta t_\text{cg} \gg \Omega^{-1}$ still leads to the FME (See Fig.\ref{f:Scenery}\textbf{c}). But the coarse-graining time can also be chosen such that $\Delta t_\text{cg} \ll \Omega^{-1}$ and Rabi oscillations are resolved. This is the case of the OBE (See Fig.\ref{f:Scenery}\textbf{d}). See also Table~\ref{t:Table} for a summary of the regimes of both descriptions. We now detail the derivation of each master equation and their regime of validity.

{\renewcommand{\arraystretch}{2.5}
\begin{table*}\begin{center}
\begin{tabularx}{18cm}{|l||Y|Y|}
\hline
& Optical Bloch Equations & Floquet Master Equation\\
\hhline{|=::=:=|}
Regime of validity& $\tau_\text{c}^{-1},\omega_\text{L},\omega_\text{qb} \gg\Omega, \gamma, \gamma \bar{n}$ & $\tau_\text{c}^{-1},\omega_\text{L},\omega_\text{qb},\Omega \gg \gamma_\text{max}$\\
\hline
Coarse-graining time $\Delta t_\text{cg}$ & $\tau_\text{c}^{-1},\omega_\text{L},\omega_\text{qb}\gg\Delta t_\text{OBE}^{-1} \gg\Omega, \gamma, \gamma \bar{n}$ & $\tau_\text{c}^{-1},\omega_\text{L},\omega_\text{qb},\Omega\gg\Delta t_\text{FME}^{-1} \gg \gamma_\text{max}$\\
\hline
\end{tabularx}
\end{center}
\caption{Regime of validity of OBE and FME and constraints on the coarse-graining time $\Delta t_\text{cg}$ used in the corresponding derivations.}\label{t:Table}
\end{table*}
}

\subsection{Floquet Master Equation (FME)}\label{s:FME}

\subsubsection{Derivation}\label{s:FMEdyn}

A sufficient -- but in general, not necessary -- solution for deriving a Lindblad equation consists in choosing $\Delta t_\text{cg}$ much larger than any oscillation frequency involved in the interaction-picture system operators $\sigma_l^I(t)$. This method was introduced in Ref.~\cite{Davies74} for systems characterized by a time-independent Hamiltonian. In the context of multipartite systems, it leads to so-called ``global'' master equations characterized by dissipation terms involving the energy spectrum of the total system \cite{Gonzalez17}. In the present case, it consists in choosing $\Delta t_\text{cg} \equiv \Delta t_\text{FME}$ (See Fig.\ref{f:Scenery}\textbf{b}, \textbf{c}) verifying 
\bb
\tau_\text{c}^{-1},\omega_\text{L},\omega_\text{qb},\Omega\gg\Delta t_\text{FME}^{-1} \gg 
\gamma_\text{max}.\label{DtFME}
\ee
Note that this choice is restricted to the strong driving case where $\Omega\gg \gamma_\text{max}$. Integrating Eq.~\eqref{eq:pre} over $t'$ erases all the terms with $l\neq l'$ and $\omega\neq\omega'$, yielding a master equation in the Linbdlad form known as Floquet Master Equation (FME). In the frame rotating at the drive frequency, and in the dressed basis $\ket{\pm}$ of the qubit, it reads $\dot{\tilde\rho}(t) = -i[\tilde H_\text{eff},\tilde\rho(t)] + \tilde{\cal L}_\text{FME}[\tilde\rho(t)]$. The Lindbladian $\tilde{\cal L}_\text{FME}= \tilde{\cal L}_0+\tilde{\cal L}_1+\tilde{\cal L}_2$,
is composed of the three following superoperators
\bb
\tilde{\cal L}_0 &=& (\gamma_{0,\downarrow}+\gamma_{0,\uparrow}){\cal D}_{\Sigma_z}\label{eq:L0}\\
\tilde{\cal L}_1 &=& \gamma_{1,\downarrow}{\cal D}_{\Sigma_-}+\gamma_{1,\uparrow}{\cal D}_{\Sigma_+}\label{eq:L1}\\
\tilde{\cal L}_2 &=& \gamma_{2,\downarrow}{\cal D}_{\Sigma_-}+\gamma_{2,\uparrow}{\cal D}_{\Sigma_+}\label{eq:L2},
\ee
which involve the Pauli matrices $\Sigma_z =\ket{+}\bra{+}-\ket{-}\bra{-}$ and $\Sigma_- =\ket{-}\bra{+} = \Sigma_+^\dagger$ in the dressed qubit basis. We denoted ${\cal D}_X[\rho] = X\rho X^\dagger - \frac{1}{2}\{ X^\dagger X,\rho\}$, with $\{A,B\} = AB+BA$, the dissipation superoperator. $\tilde{\cal L}_0$ models a pure-dephasing channel. It involves the rates:
\bb
\gamma_{0,\downarrow} &=& \dfrac{g^2}{4\Omega^2}\Gamma(\omega_\text{L})(N(\omega_\text{L})+1)\\
\gamma_{0,\uparrow} &=&\dfrac{g^2}{4\Omega^2}\Gamma(\omega_\text{L})N(\omega_\text{L}).
\ee
Conversely, $\tilde{\cal L}_1$ and $\tilde{\cal L}_2$ correspond to two thermal relaxation channels, with rates
\bb
\gamma_{1,\downarrow} &=&\frac{(\Omega+\delta)^2}{4\Omega^2}\Gamma(\omega_1)(N(\omega_1)+1)\\
\gamma_{1,\uparrow} &=&\frac{(\Omega+\delta)^2}{4\Omega^2}\Gamma(\omega_1)N(\omega_1)\\
\gamma_{2,\downarrow} &=& \frac{(\Omega-\delta)^2}{4\Omega^2}\Gamma(\omega_2)N(\omega_2)\\
\gamma_{2,\uparrow} &=&\frac{(\Omega-\delta)^2}{4\Omega^2}\Gamma(\omega_2)(N(\omega_2)+1).
\ee
We have introduced the frequencies $\omega_1 =\omega_\text{L}+\Omega$ and $\omega_2 = \omega_\text{L}-\Omega$. Together with $\omega_L$ (associated with decay channel $\tilde{\cal L}_0$), they correspond to the three frequencies of the fluorescence emission spectrum, the famous Mollow triplet \cite{Mollow69}. An intuitive explanation for these behaviors, including the  effective negative temperature of channel $\tilde{\cal L}_2$, can be obtained from the so-called radiative cascade picture, where the drive is modeled as a large coherent field injected in a single mode cavity (See \cite{CCT} and Appendix B). The FME reproduces the effective dynamics of the qubit once the field is traced out.

\subsubsection{Dynamics}

The coarse-graining time $\Delta t_\text{FME}$ yields an equation in which the dissipation captured by the Lindbladian ${\cal L}_\text{FME}$ solely involves transitions between the eigenstates of $\tilde H_\text{eff}$, i.e. the dressed qubit states. Therefore, its fixed point commutes with $\tilde H_\text{eff}$, making the dressed basis the only relevant basis of the problem. Because of this unicity of the basis, the thermodynamics of the fluorescence mechanisms captured by the FME acquires a largely classical interpretation as we show in Section~\ref{s:ThermoFME}. 
Here we highlight a few properties of the dynamics important for the thermodynamic analysis.

Firstly when expressed in the dressed basis, the FME decouples the dynamics of the populations and of the coherences, the latter decaying exponentially in time. The populations dynamics are ruled by a set of rate equations that could model a classical stochastic process, where all the transition rates are related two by two with a detailed balance condition. 

Secondly, the fixed point of the Lindbadian and the steady-state of the total master equation $\tilde\rho_\text{FME}^\infty $ are identical and read $\tilde\rho_\text{FME}^\infty = \tilde P_+^\infty \ket{+}\bra{+} + (1-\tilde P_+^\infty) \ket{-}\bra{-}$, with 
\bb
\tilde P_+^\infty = \bra{+}\tilde\rho_\text{FME}^\infty\ket{+} = \frac{\gamma_{1,\uparrow}+\gamma_{2,\uparrow}}{\gamma_{1,\uparrow}+\gamma_{1,\downarrow}+\gamma_{2,\uparrow}+\gamma_{2,\downarrow}}.\label{eq:Ppss}
\ee
As we show below, this property is essential to evidence the positivity of the entropy production rate using the same methodology as in the textbook case of slowly driven systems, which is based on the use of Spohn's equality \cite{Alicki18,Breuer}.  

These various reasons have contributed to make the FME the paradigm to investigate the thermodynamics of fluorescence \cite{Alicki13,Langemeyer14,Cuetara15,Donvil18}. However, this approach is limited since it solely captures the dynamics of the system on timescales much longer than a single Rabi period $\Omega^{-1}$, blurring out the coherent regime of light--matter interaction.

\subsection{Optical Bloch Equations (OBE)} \label{s:OBE}\label{s:DerivationOBE}

\subsubsection{Regime of validity}

The OBE were historically justified by the assumption that the presence of the drive does not affect the form of the dissipation, such that it solely depend on the bare qubit parameters \cite{CCT}. This assumption is natural for a very weak drive $g\ll \gamma(\bar n +1)$, where we have introduced $\gamma = \Gamma(\omega_\text{qb})$ and $\bar n = N(\omega_\text{qb})$. However, demonstrating its validity in the regime of strong driving requires more care and this point was debated until recently~\cite{Alicki06,Geva95}. In order to circumvent the limitations associated with the FME without performing such a seemingly crude approximation, we start from the precursor Eq.~\eqref{eq:pre} and choose a shorter coarse-graining time $\Delta t_\text{OBE}$ that fulfills (See Fig.\ref{f:Scenery}\textbf{d}):
\bb
\tau_\text{c}^{-1},\omega_\text{L},\omega_\text{qb}\gg\Delta t_\text{OBE}^{-1} \gg\Omega, \gamma(\bar n +1) \label{DtOBE}.
\ee
We first note that $\omega_\text{qb},\omega_\text{L}\gg\Omega$. This rules out the ultra-strong coupling regime between the driving field and the qubit, while still encompassing a large range of parameters from weak drive ($\Omega \ll \gamma$) to strong drive ($\Omega \gg \gamma$). Furthermore, it is required that $\tau_\text{c}^{-1}\gg \Omega \geq |\delta|, g$, a condition which can be recast in terms of the spectral density $G(\nu)$ being approximately flat on the two intervals ${\cal I}_\pm = [\pm\omega_\text{L}-\Omega,\pm\omega_\text{L}+\Omega]$, that include $\pm \omega_\text{qb}$. This condition is different from the so-called singular coupling limit \cite{Gaspard99,Esposito05}, where the correlation function of the operator $R^I(t)$ is $\delta$-correlated, which corresponds to an idealized spectral density that is flat on the whole frequency range. Though this property was used to justify the form of the OBE, it was pointed out as being in contradiction with the KMS condition, which is essential for the thermodynamic validity of the OBE \cite{Alicki06}. Conversely, the derivation presented here allows the spectral density to take two different values for positive and negative transition frequencies, i.e. for emission and absorption of photons. Namely, below we use the following approximation:
\bb
G(\nu) \to \bar{G}(\nu) = \left\{  \begin{array}{ll} G(\omega_\text{qb}) = \gamma(\bar n+1), \quad &\nu \in {\cal I}_+,\\
G(-\omega_\text{qb}) = \gamma\bar n,\quad  &\nu\in {\cal I}_-,\end{array} \right.\label{eq:flatG}
\ee
Eqs.\eqref{eq:flatG} induce a relative deviation from the KMS condition of order $\bar{G}(\nu)e^{-\hbar\nu/k_BT}/\bar{G}(-\nu) = {\cal O}(\Omega/k_\text{B}T)$ which is negligible in the regime of Ineq.~\eqref{DtOBE}. Indeed, we recall that the frequency $\tau_\text{c}^{-1}$ corresponds to the minimum of $k_BT/\hbar$ and the zero temperature bath bandwidth set by $\Gamma(\nu)$ \cite{Breuer}, so that Ineq.~\eqref{DtOBE} implies that ${\hbar\Omega\ll k_B T}$.

The above analysis reveals that the treatment of a strictly zero temperature bath is actually forbidden in the OBE approach. The regime of quantum optics that captures this situation is characterized by $\hbar\omega_\text{qb} \gg k_BT \gg \hbar\Omega$ where the thermal occupation of the bath at the qubit frequency can be neglected, $\bar{n}\sim 0$. This textbook case is studied in \cite{Carmichael1} section 2.3.2. Below we extend the study to the case where $\bar{n}$ can be non-negligible.\\

\subsubsection{Derivation and dynamics}
The inequality $\omega_\text{L}\Delta t_\text{OBE} \gg 1$ allows us to neglect  the terms with $l\neq l'$ in Eq.~\eqref{eq:pre}, while the terms $\omega\neq\omega'$ remain. Such step corresponds to a partial secular approximation, that was recently shown to lead to accurate master equations in the case of multi-partite systems \cite{Cresser17,Cattaneo19}. We obtain:
\begin{widetext}
\bb
\dot\rho^I\!(t) &=&  \dfrac{1}{2\Delta t_\text{cg}}\int_t^{t+\Delta t_\text{cg}}\!\!\! dt' \sum_{\omega} e^{i \omega t'}G(-\omega_\text{L}+\omega)\!\Big(\sigma_{+}(\omega)\rho^I(t) \sigma^I_{-}(t') -
 \sigma^I_{-}(t')\sigma_{+}(\omega)\rho^I\!(t)  \Big) \nonumber\\
 &&+\dfrac{1}{2\Delta t_\text{cg}}\int_t^{t+\Delta t_\text{cg}}\!\!\! dt' \sum_{\omega} e^{i \omega t'}G(\omega_\text{L}-\omega)\!\Big(\sigma_{-}(\omega)\rho^I(t)\sigma^I_{+} (t') -
\sigma^I_{+} (t')\sigma_{-}(\omega)\rho^I\!(t)  \Big) \!+\! \text{H.c.}\label{eq:GBE}
\ee
\end{widetext}
We have made use of the identity $\sum_{\omega'}\sigma_{l}(\omega')e^{i\omega't'}e^{il\omega_\text{L}t'}=\sigma^I_{l}(t')$. If one stops the derivation at this point and write Eq.~\eqref{eq:GBE} in the Schr\"odinger picture, one obtains the Generalized Bloch Equations (GBE) as introduced in \cite{Geva95}. These equations can be written in a Lindblad form, and allow us to describe the evolution of the qubit over short time-scales, while still taking into account the local variations of the bath spectral density, e.g. due to the differences of thermal occupations $N(\omega_\text{L}-\omega)$ for the three different values of $\omega$. The resulting equation is still global as it involves the spectral density of the bath evaluated at all the transition frequencies of Hamiltonian $\tilde H_\text{eff}$. Using Eq.~\eqref{eq:flatG} and Eq.~\eqref{eq:GBE}, and summing over $\omega$ yields

\bb
\dot\rho^I\!(t) &=&  \dfrac{\gamma \bar n}{\Delta t_\text{OBE}}\!\int_t^{t+\Delta t_\text{OBE}}\!\!\!\!\!  \!\!\!\!\! dt'  {\cal D}_{\sigma_+^I(t')}[\rho^I(t)]\nonumber\\
&& +\dfrac{\gamma (\bar n+1)}{\Delta t_\text{OBE}}\!\int_t^{t+\Delta t_\text{OBE}}\!\!\!\!\!  \!\!\!\!\! dt'  {\cal D}_{\sigma_-^I(t')}[\rho^I(t)].\label{eq:OBE0}
\ee
Once rewritten in Schr\"odinger picture and after performing the integration over $t'$ (see e.g. Section IV.B.4 of Ref.~\cite{CCT} for a detailed explanation of these derivation steps), Eq.~\eqref{eq:OBE0} finally takes the well-known form of the OBE, which in the rotating frame read
\bb
\dot{\tilde\rho}(t) = -\dfrac{i}{\hbar}[\tilde H_\text{eff},\tilde\rho(t)] + {\cal L}_\text{OBE}[\tilde\rho(t)] \label{eq:OBErot}.
\ee
\noindent ${\cal L}_\text{OBE}$ is the Lindbladian super-operator accounting for the non-unitary part of the dynamics:
\bb
{\cal L}_\text{OBE}[\rho] = \gamma \bar{n} {\cal D}_{\sigma_+}[\rho] + \gamma(\bar{n}+1) {\cal D}_{\sigma_-}[\rho] \label{eq:LindbladianOBE}.
\ee
From Eq.~\eqref{eq:OBErot}, one can note a crucial difference with the FME. Namely, the fixed point of ${\cal L}_\text{OBE}[\tilde\rho(t)]$ does not commute with $\tilde H_\text{eff}$. This situation gives rise to a genuinely quantum non-equilibrium dynamics. More precisely, in the bare qubit basis $\{\ket{0},\ket{1}\}$, the dissipation exhibits the structure of a rate equation which does not couple populations and coherences. But in this basis, the Hamiltonian $\tilde H_\text{eff}$ is not diagonal and induces coherences. Therefore, in contrast with the FME, the OBE do not take the form of a rate equation neither in the bare qubit basis $\{\ket{0},\ket{1}\}$, nor in the eigenbasis of $\tilde H_\text{eff}$ and cannot be interpreted as a classical stochastic process.
The consequences of this property on the thermodynamic behavior are analyzed in Section \ref{s:Signatures}.

\subsection{FME versus OBE}\label{s:FMEvsOBE}

We now focus on the regime $\tau_\text{c}^{-1} \gg \Omega \gg \gamma$, where the two choices of coarse-graining can be made and both the FME and OBE descriptions are valid. In practice in an experiment, the time resolution of the measuring apparatuses used to probe the dynamics, that we denote $\tau_\text{meas}$, determines the relevant description. If $\tau_\text{meas}$ is shorter than the Rabi period $\Omega^{-1}$, the OBE description will provide the most detailed description. If instead $\tau_\text{meas} \gtrsim \Omega^{-1}$, the FME description is relevant. It is in this case equivalent to the OBE coarse-grained over the time record of the apparatus, as we now proceed to show.
We start from the OBE in the interaction picture with respect to $\tilde H_\text{eff}$:
\bb
\dot\rho^I(t) = \gamma( \bar{n}+1){\cal D}_{\sigma_-^I(t)}[\rho^I(t)] +  \gamma \bar{n}{\cal D}_{\sigma_-^I(t)}[\rho^I(t)].\nonumber\\
\ee
Coarse-grain this master equation choosing the time-scale leading to the FME, i.e. $\Delta t_\text{FME}$ fulfilling Ineq.~\eqref{DtFME}, we obtain:
\begin{flalign}
\rho^I(t+\Delta t_\text{FME})-\rho^I(t) &=\nonumber \\ \int_t^{t+\Delta t_\text{FME}} dt'\bigg(\gamma(\bar{n}+&1){\cal D}_{\sigma_-^I(t')} +  \gamma \bar{n}{\cal D}_{\sigma_+^I(t')}\bigg)\rho^I(t').\label{D:pre}\nonumber\\
\end{flalign}
As in Section~\ref{s:Redfield}, the evolution of the qubit density matrix in the interaction picture $\rho^I(t)$ can be neglected over the time-scale defined by $\Delta t_\text{FME}$. We therefore replace $\rho^I(t')$ with $\rho^I(t)$ in the integrand of Eq.~\eqref{D:pre}. We then use the Fourier decomposition of $\sigma^I_\pm(t)$ (see Eqs.~\eqref{d:sigpm0}-\eqref{d:sigpmp}), and expand the terms ${\cal D}_{\sigma_\pm^I(t')}$ of the integrand of Eq.~\eqref{D:pre}. Written in the dressed basis, we gather the terms proportional to ${\cal D}_{\Sigma_\pm^I(t')}$ and ${\cal D}_{\Sigma_z^I(t')}$, and other ``non-diagonal'' terms featuring oscillating coefficients $e^{\pm i\Omega t'}$ or $e^{\pm 2i\Omega t'}$. Such fast oscillations average to zero over $\Delta t_\text{FME}\gg \Omega^{-1}$. Neglecting these terms (which is equivalent to the Secular Approximation needed to derive the FME), 
we find eventually:
\bb
{\dot \rho}^I(t) &=& \gamma\left[\frac{(\Omega+\delta)^2}{4\Omega^2}(\bar{n}+1) + \frac{(\Omega-\delta)^2}{4\Omega^2}\bar n\right]{\cal D}_{\Sigma_-}\nonumber\\
&&+\gamma\left[\frac{(\Omega-\delta)^2}{4\Omega^2}(\bar{n}+1) + \frac{(\Omega+\delta)^2}{4\Omega^2}\bar n\right]{\cal D}_{\Sigma_+}\nonumber\\
&&+\gamma\dfrac{g^2}{4\Omega^2}(2\bar{n} + 1){\cal D}_{\Sigma_z}\label{eq3:FMEflat}.
\ee
Eq.~\eqref{eq3:FMEflat} corresponds to the FME if we make the approximations $\Gamma(\omega_j) \simeq \gamma$, $N(\vert\omega_j\vert)\simeq \bar n$. These approximations are legitimate in the common regime of validity of the OBE and the FME. 

The derivation above shows that in this regime the FME is a time-average of the OBE. Therefore, it does not yield more information about the dynamics. On the contrary, this coarse-graining averages out quantum properties of the dynamics that are captured by the OBE, such as the coherent nature of the energy exchange between the driving field and the qubit. A similar procedure would allow us to derive the FME from the GBE (Eq.~\eqref{eq:GBE}), leading to a master equation similar to Eq.~\eqref{eq3:FMEflat}, but which takes into account the frequency dependence of the mode thermal occupation. Once again, the FME does not bring more information than the GBE about the qubit dynamics in the range of parameters where both are valid. Finally, in the present context, the FME turns out to be advantageous solely in the case where the bath correlation time is of the order or larger than a single Rabi oscillation, for instance in the presence of a structured bath \cite{Lewenstein87,Murch12}, or for extremely low temperatures $T\lesssim \hbar\Omega/k_B$. The various conditions of validity of FME and OBE are summarized in Table \ref{t:Table}.

\begin{figure*}
\begin{center}
\includegraphics[width=0.9\textwidth]{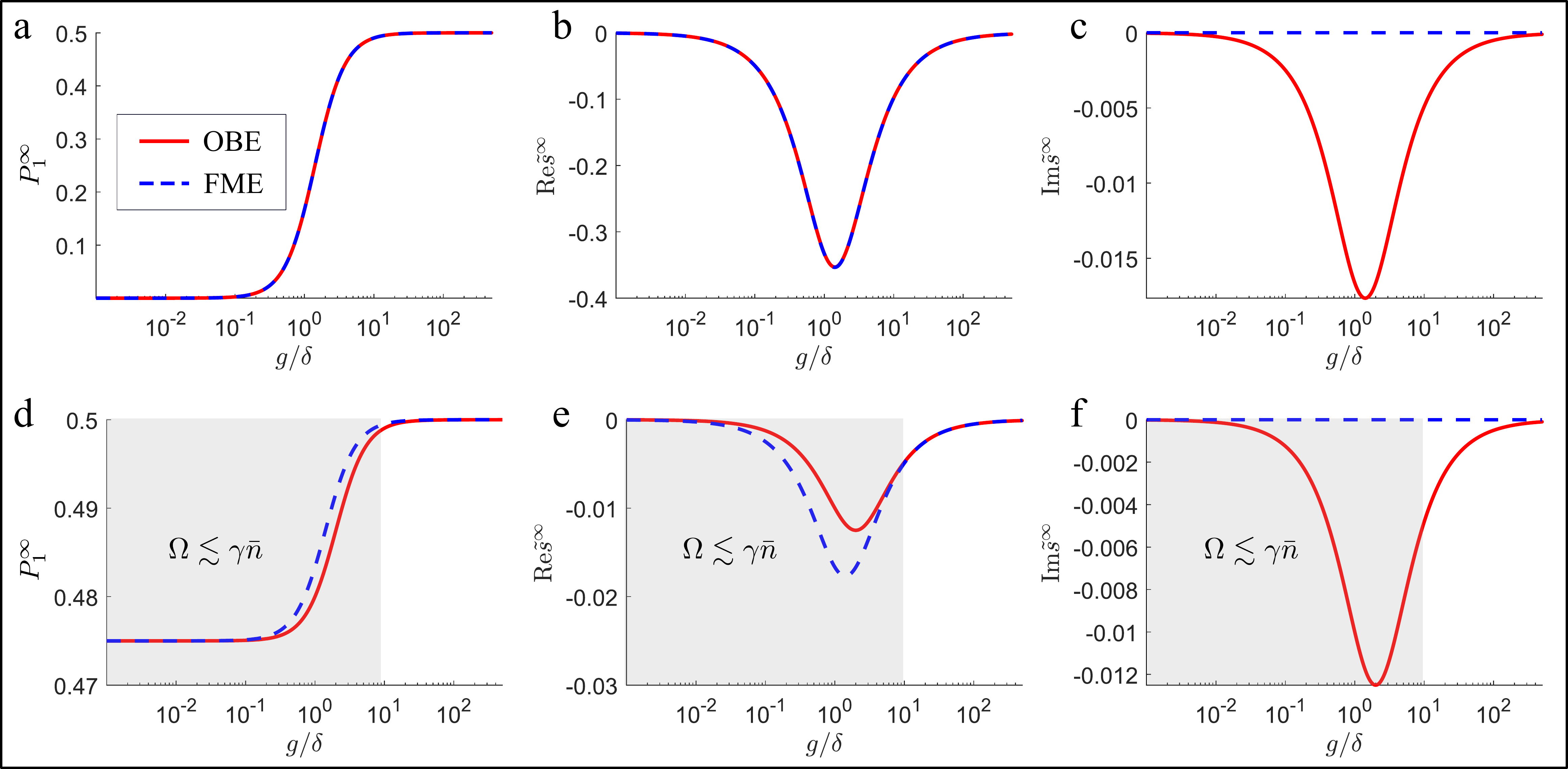}
\end{center}
\caption{Comparison of steady-state values of the population $P_1^\infty = \bra{1}\tilde\rho^\infty\ket{1}$ (\textbf{a},\textbf{d}) and the real (\textbf{b},\textbf{e}) and imaginary (\textbf{c},\textbf{f}) part of the amplitude of the coherences $\tilde s= \bra{1}\tilde\rho^\infty\ket{0}$ in the bare qubit basis $\{\ket{0},\ket{1}\}$, in the rotating frame, for the OBE (red solid) and FME (blue dashed). The analytic expression of the OBE steady-state is given in Appendix C. Parameters: $\delta/\omega_\text{L} = \numprint{1E-3}$, $\gamma/\omega_\text{L} = \numprint{1E-4}$, $\hbar\omega_\text{L}/k_BT = 10$ (\textbf{a}, \textbf{b}, \textbf{c}), $\hbar\omega_\text{L}/k_BT = 0.1$ (\textbf{d}, \textbf{e}, \textbf{f}). The grey area corresponds to values of $g/\delta$ such that $\Omega \lesssim \gamma\bar n$ and therefore the FME is expected not to be valid anymore.
\label{f:Comp}
}
\end{figure*}


We have computed the expressions of the main dynamical quantities corresponding to the steady-state of the FME and the OBE in their common regime of validity (see Appendix C). The populations and coherences in the basis $\{\ket{1},\ket{0}\}$ are plotted in Fig.~\ref{f:Comp} showing a very good agreement in the common domain of validity. The deviations between the two predicted steady-states (and in particular the imaginary part of coherences which is non zero for OBE and strictly zero for FME) vanish in the limit $\gamma/g \to 0$, $\hbar\Omega/k_\text{B}T \to 0$ and $\vert\delta\vert/\omega_\text{L} \to 0$. These are the three conditions assumed in the derivation of either the FME or the OBE (see Appendix C). Finally, we observe deviations between the two approaches at large temperature such that $\bar n>1$. These deviations can be explained noting that in this case $\gamma_\text{max}\sim \gamma\bar n \gg \gamma$ is the typical evolution rate of the qubit density operator in the interaction picture for high temperatures (rather than $\gamma$), and therefore the validity of the derivation of the FME requires $\Omega \gg \gamma\bar n$ (see Eq.~\eqref{DtFME}). While this fact does not affect the global agreement of both methods within the common regime of validity, it points out the interest of the OBE to describe moderate drives such that $\Omega \lesssim \gamma$, or $\Omega \lesssim \gamma\bar n$ which cannot be captured by the FME. We have also compared the predictions of both approaches in the very low temperature regime where $k_\text{B}T \lesssim \hbar\Omega$. The OBE are expected to break down in this regime due to the rapid variation of the environment spectral density $G(\nu)$ in the vicinity of $\omega_\text{L}$. Suprisingly, the OBE turn out to be very robust and no appreciable difference can be found numerically with the steady state predicted by the FME (which are valid in this regime). We attribute this phenomenon to the fact that, although the spectral density is varying rapidly as a function of $\nu$, the amplitude of this variation is proportional to $\bar n$ which vanishes exponentially fast as $\hbar\omega_\text{L}/k_\text{B}T$ increases, yielding a negligible effect on the qubit dynamics.


\section{Thermodynamic analysis}
\label{s:Thermo}

In the previous section, we have provided rigorous derivations of the OBE and the FME, and unambiguously clarified how their respective regimes of validity depend on the microscopic parameters of the bath. In the rest of the paper, we apply a similar methodology to derive consistent expressions of the First and Second Laws of thermodynamics respectively associated to each description. After drawing a brief state of the art, we present our general strategy to derive the heat, work and entropy flows (\ref{s:GeneralSt}). We then successively apply this strategy to the case of the FME (\ref{s:ThermoFME}), then the OBE (\ref{s:ThermoOBE}).

\subsection{General strategy}\label{s:GeneralSt}

\subsubsection{State of the art}\label{s:ThermoSOA}

The thermodynamic description of a driven quantum system ${\cal S}$ weakly coupled to a thermal bath is well-known \cite{Alicki18,Lebowitz78,Alicki79} if the following conditions are satisfied:
(i) The drive is ``adiabatic", which means that the typical time-scale of variation of the system's Hamiltonian $H_{\cal S}(t)$ is much longer than the coarse-graining time $\Delta t_\text{cg}$ used to derive the master equation, and (ii) $\Delta t_\text{cg}$ is long enough to resolve all the transition frequencies in the spectrum of $H_{\cal S}(t)$. The master equation then takes the form ${\dot \rho_{\cal S}(t) = -i [H_{\cal S}(t),\rho] +{\cal L}_t[\rho_{\cal S}]}$, where the Lindbladian ${\cal L}_t$ adapts every time-step $\Delta t_\text{cg}$ to the Hamiltonian $H_{\cal S}(t)$. Pioneer studies performed in this regime have found the flow of work $\dot W_\text{adiab}(t)$ provided by the drive, and heat $ \dot Q_\text{adiab}(t)$ exchanged with the thermal bath to be \cite{Lebowitz78,Alicki79}:
 \bb
 \dot W_\text{adiab}(t) &=& \text{Tr}\{\rho_{\cal S}(t){d_t H}_{\cal S}(t)\}\label{eq:Wad}\\
 \dot Q_\text{adiab}(t) &=& \text{Tr}\{{\cal L}_t[\rho_{\cal S}(t)]H_{\cal S}(t)\}.\label{eq:Qad}
 \ee
Note that above and in the remainder of this article, when dealing with variations of thermodynamic quantities, we use the notation $d_t$ for time-derivative of state-variables, like the internal energy $U(t) = \text{Tr}\{\rho(t)H(t)\}$ or the Von Neumann entropy $S(t) = -k_\text{B}\text{Tr}\{\rho(t)\log\rho(t)\}$ of the qubit, and the dot for the rates associated with path-dependent variables like the heat, work or entropy production. 
The expressions for the work and heat flows reported in Eq.~\eqref{eq:Wad} and \eqref{eq:Qad} are validated by their consistency with the Second Law, which boils down to proving the positivity of the entropy production rate ${\dot \sigma_\text{adiab.} = d_t S - \dot{Q} _\text{adiab}(t)/T}$. This proof is based on Spohn's inequality \cite{Spohn78,Breuer} ${\dot\sigma_\text{adiab.} = -k_\text{B}\text{Tr}\{{\cal L}_t[\rho_{\cal S}(t)](\log\rho_{\cal S}(t) - \log \pi_{\cal S}(t))\}\geq 0}$, where ${\pi_{\cal S}(t) = e^{-H_{\cal S}(t)/k_\text{B}T}/Z_{\cal S}(t)}$ is the instantaneous fixed point of ${\cal L}_t$, with $Z_{\cal S}(t)$ the corresponding instantaneous partition function. 
  
Conversely in the situation described by the OBE and the FME, the characteristic time of variation of $H_{\cal S}(t)$ is $\omega_\text{L}^{-1}$. It is thus much shorter than $\Delta t_\text{cg}$, as required by the Secular Approximation (see Section \ref{s:ME}). Therefore the usual methodology cannot be applied directly. It was already shown in some cases going beyond the assumptions of Ref.~\cite{Alicki79} that the correct splitting between work and heat, i.e. which is in agreement with the Second Law, requires more care. For instance, in the case of local Lindblad equations, one may need to specify the underlying microscopic model \cite{DeChiara2018Nov}. Moreover, studies based on the FME have used different techniques to infer \cite{Alicki13,Donvil18} or derive \cite{Langemeyer14,Cuetara15} the heat flow and check its compatibility with the Second Law, and the found expression differs notably from the adiabatic expression Eq.~\eqref{eq:Qad}.

This apparent flexibility in the definitions of the work and heat flows is a consequence of the fact that open system's dynamical equations like the FME or the OBE do not contain information about the dynamics of the bath and can in principle originate from several microscopic models, each of them potentially associated with different thermodynamic balances. 
To solve that issue and derive expressions for the thermodynamic quantities that correspond to the situation we actually want to describe, we need to start from a global level of description that incorporates the bath. In this way, we can define and compute the heat flow as the energy provided by the thermal bath to the driven qubit. Beside being natural, this choice of definition is compliant with the Second Law of thermodynamics, provided the system and bath evolve through a global unitary \cite{Esposito10Entropy}. 

\subsubsection{Definition of the heat and proof of the Second Law} \label{s:NJP}

In this section we apply to our model the strategy reported in Ref.~\cite{Esposito10Entropy}. Namely, we provide a thermodynamic analysis of the joint qubit-bath system while it evolves unitarily between $t$ and $t+\Delta t_\text{cg}$.  At time $t$, the joint qubit-bath system is assumed to be in a product state $\rho_{\cal SB}(t) = \rho(t)\otimes\rho_{\cal B}^\text{eq}$, and evolves into $\rho_{\cal SB}(t+\Delta t_\text{cg}) = {\cal V}_{\Delta t_\text{cg}}\rho(t)\otimes\rho_{\cal B}^\text{eq}{\cal V}_{\Delta t_\text{cg}}^\dagger$ where ${\cal V}_{\Delta t_\text{cg}}$ is the operator ruling the unitary evolution. Conversely, the reduced state of the qubit $\rho(t+\Delta t_\text{cg}) = \text{Tr}_{\cal B}[\rho_{\cal SB}(t+\Delta t_\text{cg})]$ can be derived from the OBE or the FME, depending on the regime of parameters and the chosen coarse-graining time. We finally define the heat increment between $t$ and $t+\Delta t_\text{cg}$ as the opposite of the bath energy change:
\bb
\Delta t_\text{cg} \dot Q(t) &=&  - \Delta t_\text{cg} d_t E_{\cal B}(t)\nonumber\\
&=& \text{Tr}\{H_{\cal B}(\rho_{\cal RS}(t)-\rho_{\cal RS}(t+\Delta t_\text{cg}))\}. \label{defHeatGen}
\ee
Interestingly, Eq.~\eqref{defHeatGen} bridges a gap between two pictures. On the one hand, the system-bath evolution is unitary, such that the bath energy change corresponds to the variation of a state-variable and is denoted by $d_t$. On the other hand, the system evolution is ruled by a Lindblad master equation where no information on the bath state is available. In this case the heat flow becomes a path-dependent variable labelled by a dot.

A natural expression for the entropy production $\dot \sigma(t)$ is:
\bb
\Delta t_\text{cg} \dot \sigma(t) = \Delta t_\text{cg}\left( d_t S(t) + \frac{d_t E_{\cal B}(t)}{T}\right).\label{eq:ep1}
\ee
The whole point of the proof is to evidence the positivity of this rate. This is done by rewriting the right-hand side of Eq.~\eqref{eq:ep1} as a quantum relative entropy $D(\rho_1\vert\vert\rho_2) = \text{Tr}\{\rho_1(\log\rho_1-\log\rho_2)\}$, whose positivity is ensured. We have:
\bb
&&D\left(\rho_{\cal SB}(t+\Delta t_\text{cg}) \vert\vert \rho(t+\Delta t_\text{cg})\otimes \rho_{\cal B}^\text{eq}\right)\nonumber\\  &=& \text{Tr}\left\{\rho_{\cal SB}(t+\Delta t_\text{cg})\log \rho_{\cal SB}(t+\Delta t_\text{cg}\right\} \nonumber\\
&&- \text{Tr}\left\{\rho_{\cal SB}(t+\Delta t_\text{cg}) \log(\rho(t+\Delta t_\text{cg})\otimes \rho_{\cal B}^\text{eq})\right\}\nonumber\\
 &=& -S(t)-S_{\cal B}^\text{eq} + S(t+\Delta t_\text{cg})\nonumber\\
 && + \text{Tr}\left\{\rho_{\cal SB}(t+\Delta t_\text{cg})\left( \frac{H_{\cal B}}{T}+ k_\text{B}\log Z_{\cal B}\right)\right\}\nonumber\\
 &=& \Delta t_\text{cg} \left(d_t S(t)+ \frac{d_t E_{\cal B}(t)}{T}\right).
\ee
To obtain the second equality, we used that the Von Neumann entropy of $\rho_{\cal SB}$ is conserved during the unitary ${\cal V}_{\Delta t_\text{cg}}$, and therefore its value at $t+\Delta t_\text{cg}$ is equal to its initial value $S(t)+S_{\cal B}^\text{eq}$ where we have denoted $S_{\cal B}^\text{eq} = -k_\text{B}\text{Tr}\{\rho_{\cal B}^\text{eq}\log\rho_{\cal B}^\text{eq}\}$ the Von Neumann entropy of the bath at equilibrium. The last equality is obtained using the expression of the entropy of the bath after the interaction with the system, namely $S_{\cal B}(t+\Delta t_\text{cg}) = E_{\cal B}(t+\Delta t_\text{cg})/T + k_\text{B}\log Z_{\cal B}$. 

This derivation is general in the sense we did not have to specify the details of the evolution. It will therefore be valid for both the OBE and the FME, provided we are able to compute in each case the energy change of the bath.

\subsubsection{Heat flow exchanged with the bath}

The next step of our method is to compute the energy change of the heat bath within the regimes leading to the OBE and the FME. To do so, we re-start from Eq.~\eqref{eq:Redfield} to formally express the coarse-grained energy variation of the bath:\bb
&&d_t E_{\cal B}(t) = -\frac{1}{\hbar^2\Delta t_\text{cg}}\!\displaystyle\int_{t}^{t+\Delta t_\text{cg}}\!\!\!\!\!\!\!\! dt'\!\!\int_{t}^{t'}\!\!\!\! dt''\nonumber\\
&&\quad\quad\text{Tr}_{\cal B}\!\Big\{H_{\cal B}\left[H^I_{\cal RS}(t'),\!\left[H^I_{\cal RS}(t''),\rho^I(t)\!\otimes\!\rho^I_{\cal B}\right]\right]\!\!\Big\}.\quad\quad
\label{eq:RedfieldER}
\ee
We then apply to Eq.~\eqref{eq:RedfieldER} the same procedure used to derive the FME and OBE. Namely, we change the variable $t''$ to $\tau = t'-t''$, extend the upper limit of the integral over $\tau$ to $+\infty$ and use the explicit form of $H_{\cal SB}^I(t)$. We get

\bb
&&d_t E_{\cal B}(t)=  \dfrac{1}{4\Delta t_\text{cg}}\int_t^{t+\Delta t_\text{cg}}\!\!\!\!\!\! dt' \sum_{ll'} \sum_{\omega,\omega'} e^{i(\omega- \omega')t'+i(l-l')\omega_\text{L}t'}\nonumber\\
&&\quad\quad\quad\quad\quad\quad\times A(-\omega - l\omega_\text{L})\moy{\tilde\sigma_{l'}^\dagger(\omega')\tilde\sigma_{l}(\omega)}+\text{c.c.},\label{eq:preER}
\ee

where we introduced 
\bb
{A(\nu) = (1/\hbar^2)\int_{-\infty}^\infty d\tau e^{i\nu\tau} \moy{[R^I(\tau),H_{\cal B}]R^I(0)}}.
\ee
The imaginary part of $\int_{0}^\infty d\tau e^{i\nu\tau} \moy{[R^I(0),H_{\cal B}]R^I(\tau)}$ is a contribution of similar order and form as the Lamb shift. We therefore neglect it to be consistent with the dynamical description. We then insert the explicit expression of $R$ and $H_{\cal B}$ to obtain
\bb
A(\nu) = \hbar\nu G(\nu).\label{eq:Anu}
\ee
The last step of the derivation is the Secular Approximation. As for the dynamics, two choices are possible, involving respectively the coarse-graining times $\Delta t_\text{FME}$ and $\Delta t_\text{OBE}$ introduced in Section \ref{s:Redfield}. As we will now see, each of these choices leads to a different expression of the heat flow. This gives rise to two different expressions of the First and Second Laws of thermodynamics, consistent with the FME and the OBE, respectively.

\subsection{Thermodynamics of the FME}\label{s:ThermoFME}

 As a consistency check, we first compare our results to former thermodynamic studies of FME (See ~\cite{Alicki13,Langemeyer14,Cuetara15,Donvil18}). While we obtain identical expressions for the heat flow, our framework sheds new light on the work mechanism at play, which has the particularity to be non-unitary and thus drastically different from usual work exchanges in quantum thermodynamics. 

\subsubsection{First and Second Laws}

Choosing $\Delta t_\text{cg} \equiv \Delta t_\text{FME}$ verifying Eq.~\eqref{DtFME} leads to neglect the terms with $l\neq l'$ and $\omega\neq\omega'$ in Eq.~\eqref{eq:preER}, yielding:
\bb
&&d_t E_{\cal B}(t)=  \dfrac{1}{4} \sum_{l,\omega}A(-\omega - l\omega_\text{L})\moy{\sigma_{l}^\dagger(\omega)\sigma_{l}(\omega)}+\text{c.c.}\quad\quad
\ee
Inserting Eq.~\eqref{eq:Anu} and defining the heat flow as $\dot Q_\text{FME} = - d_t E_{\cal B}(t)$, we finally obtain:
\bb
\dot Q_\text{FME} &=& - \sum_{j=1}^2 \hbar\omega_j \left( \gamma_{j,\downarrow}P_+(t)-\gamma_{j,\uparrow}P_-(t)\right)\nonumber\\ 
&&- \hbar\omega_\text{L}(\gamma_{0,\downarrow}-\gamma_{0,\uparrow}).
\label{eq:QFME}
\ee
This expression matches the heat flow obtained in Refs.~\cite{Alicki13,Langemeyer14,Cuetara15,Donvil18}. Using that the energy variation of the joint qubit-bath system solely comes from the driving, we deduce the work flow $\dot W_\text{FME}(t)= d_t E_{\cal B}(t) + d_t U(t)$. It verifies:
\bb
\dot W_\text{FME} &=& \hbar\omega_\text{L}\Big((\gamma_{1,\downarrow}-\gamma_{2,\downarrow})P_+(t)- (\gamma_{1,\uparrow}-\gamma_{2,\uparrow})P_-(t)\nonumber\\
&&+(\gamma_{0,\downarrow}-\gamma_{0,\uparrow})\Big).\label{eq:WFME}
\ee
The First Law of thermodynamics then reads $d_t U(t) = \dot W_\text{FME} + \dot Q_\text{FME}$. Focusing on the Second Law, the entropy production rate reads:
\bb
\dot\sigma_\text{FME}(t) &=& d_t S(t) - \frac{\dot Q_\text{FME}(t)}{T}\nonumber\\
&=& - \sum_{j=1}^{2} \text{Tr}\left\{\tilde{\cal L}_j[\tilde\rho(t)]\left(\log\tilde\rho(t)-\log\tilde\pi_j\right)\right\}\nonumber\\
&& + \frac{\hbar\omega_\text{L}}{T}\frac{g^2}{4\Omega^2}\Gamma(\omega_\text{L})\geq 0.\label{eq:sigmaFME}
\ee 
We introduced the fixed points $\tilde\pi_j = \exp(-\hbar\omega_j \Sigma_z /k_BT)/Z_j$ of the dissipation superoperators $\tilde{\cal L}_j$ for $j=1,2$ where $Z_j$ are the corresponding partition functions, and $\Sigma_z = \ket{+}\bra{+}-\ket{-}\bra{-}$ the population inversion in the dressed basis. The positivity of $\dot \sigma$ is ensured by the general demonstration presented above. It is interesting to notice that in the specific case of the FME, this positivity also appears as a consequence of Spohn inequality \cite{Spohn78,Alicki79,Breuer}, just as for the adiabatic master equations.

\subsubsection{Non-unitary work exchange}

The expressions Eqs.~\eqref{eq:QFME} and \eqref{eq:WFME} are very different from their counterparts in the case of an adiabatic driving (see above and \cite{Lebowitz78,Alicki79}). Remarkably, the expression of the work flow involves dissipative rates that characterize the action of the bath. The interpretation of this hybrid behavior is clear when recalling that the FME describes transitions induced by the bath and assisted by the driving field which can take or provide a quantum of energy $\hbar\omega_\text{L}$ (see Appendix B). In this picture, the energy provided by the driving translates as the work of a non-conservative force which cannot be expressed as an energy change of the system, but is rather immediately converted into heat. Such forces generically appear when a degree of freedom contributing to the total energy change between states is not explicitly described but must nevertheless be accounted for \cite{Esposito15,Raz16,Fogedby17,Rao18}. In the FME description, the photons in the driving field are not accounted for in the system's energy even though each transition induced by the bath is actually associated with a variation of this photon number. 

We note that the exchanges of photons between the qubit and driving field can be explicitly included as in the radiative cascade model of the driven qubit described in Appendix B. This model provides a good interpretation for the constant contributions in the heat flow and the entropy production (last lines of Eqs.~\eqref{eq:QFME} and \eqref{eq:sigmaFME}).

\subsection{Thermodynamics of the OBE}\label{s:ThermoOBE}

We now choose $\Delta t_\text{cg}= \Delta t_\text{OBE}$ and neglect the terms with $l\neq l'$ in Eq.~\eqref{eq:preER} while keeping the terms $\omega\neq\omega'$. We also exploit the flat spectrum assumption already used to derive the OBE, which amounts to the replacement $A(-\omega - l\omega_\text{L}) \simeq  -l(\omega+l\omega_\text{L})G(- l\omega_\text{L})$. Computing the integration over $t'$ up to first order in $\Omega\Delta t_\text{cg} \ll 1$ leads to the final expression:
\bb
&&d_t E_{\cal B}(t)=  \sum_{\omega,l} \hbar(\omega - l\omega_\text{L})G(-l\omega_\text{L})\text{Re}\moy{\tilde\sigma_{l}^\dagger\tilde\sigma_{l}(\omega)}\nonumber\\
&&= \gamma \sum_{\omega} \hbar(\omega +\omega_\text{L})(\bar n+1) \text{Re}\moy{\tilde\sigma_{+}\tilde\sigma_{-}(\omega)}+\nonumber\\
&&\quad\quad - \gamma\sum_{\omega} \hbar(\omega -\omega_\text{L})\bar n \text{Re}\moy{\tilde\sigma_{-}\tilde\sigma_{+}(\omega)}\nonumber\\
&&= \gamma \hbar\omega_\text{qb}\Big((\bar n+1) P_1(t)-\bar nP_0(t)\Big) + \gamma \hbar g\frac{2\bar n +1}{2} \text{Re}\tilde s(t).\nonumber
\label{eq:ER}
\ee

It is then straightforward to prove that
\bb
\dot Q(t) &=& - d_t E_{\cal B}(t)\nonumber\\
&=& \text{Tr}\{H(t){\cal L}_\text{OBE}[\rho(t)]\}.\label{eq:QER}
\ee
As before, we use energy conservation between the bath and the qubit to obtain the expression of the work flow:
\bb
\dot W(t) &=& -\hbar\omega_\text{L}g\text{Im}\tilde s(t)\label{eq:dWOBE}\nonumber\\
&=& \text{Tr}\{\rho(t)d_t H(t)\},
\ee
The First Law finally reads $d_t U(t) = \dot{W} + \dot{Q}$. Remarkably, and in sharp contrast with the case of the FME, the work and heat flows in the regime captured by the OBE have expressions similar to the case of an adiabatic driving (see above and \cite{Lebowitz78,Alicki79}). Note that when computing the corresponding expressions for the FME state and dynamics, e.g. $\text{Tr}\{H(t){\cal L}_\text{OBE}[\rho(t)]\}$ or $\text{Tr}\{\rho(t)d_t H(t)\}$, one does not get a correct estimate of the heat and work flows in the FME case. A striking evidence is that the steady state value of these quantities is both $0$, whereas Eqs.~\eqref{eq:QFME}-\eqref{eq:WFME} and \eqref{eq:QER}-\eqref{eq:dWOBE} admit a non-zero steady state values.

\begin{figure}
\begin{center}
\includegraphics[width=0.33\textwidth]{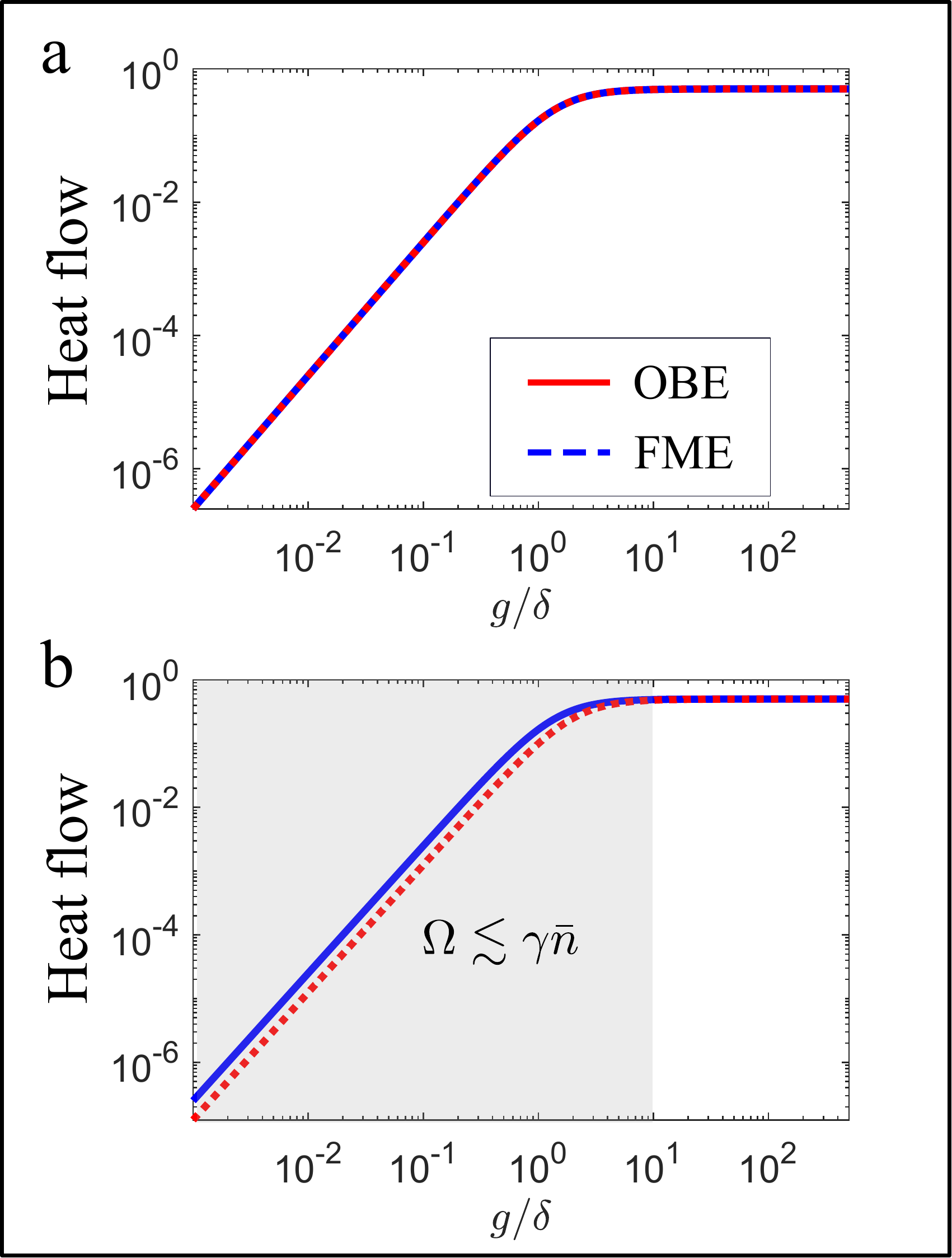}
\end{center}
\caption{Comparison of steady-state values of the heat flow for the OBE (red solid) and FME (blue dashed). Parameters: $\delta/\omega_\text{L} = \numprint{1E-3}$, $\gamma/\omega_\text{L} = \numprint{1E-4}$, $\hbar\omega_\text{L}/k_BT = 10$ (\textbf{a}), $\hbar\omega_\text{L}/k_BT = 0.1$ (\textbf{b}). The grey area corresponds to values of $g/\delta$ such that $\Omega \lesssim \gamma\bar n$ where the FME regime of validity breaks down.
\label{f:CompThermo}
}
\end{figure}

As far as the Second Law is concerned, and contrary to the adiabatic case, one cannot link $\dot Q(t)$ to the fixed point $\pi = e^{-H_\text{qb}/k_\text{B}T}/Z$ of the Lindbladian ${\cal L}_\text{OBE}$, i.e. $\dot Q(t) \neq - T\text{Tr}\{{\cal L}_\text{OBE}[\rho(t)]\log \pi\}$. Consequently, Spohn's inequality cannot be used to prove the positivity of the entropy production rate. However, the result of section \ref{s:NJP} applies and ensures that Eq.~\eqref{eq:entprod} satisfies the Second Law, namely
\bb
\dot \sigma = d_t S(t) - \frac{\dot Q(t)}{T} \geq 0. \label{eq:entprod}
\ee
As a final verification of the consistency of our results, we compare the steady-state heat and work flow derived for the OBE from those obtained in Ref.~\cite{Alicki13,Langemeyer14,Donvil18} working with the FME, given by Eqs.~\eqref{eq:QFME}-\eqref{eq:WFME}. As above, we focus on the regime where both the OBE and the FME are valid descriptions, as captured by Eq.~\eqref{eq:CommonRegime}. Fig.~\ref{f:CompThermo} shows the excellent agreement between the two approaches.

\section{Quantum thermodynamic signatures}
\label{s:Signatures}

In the Sections above, we have established a consistent thermodynamic framework for the OBE, that enabled us to capture the coherent regime of the qubit-light interaction. In this final Section, we exploit this framework to single out ``quantum thermodynamic signatures" in the mechanism of resonance fluorescence. Such signatures represent new components in heat, work and entropy flows, that are related to the presence of quantum coherences in the driven qubit. Identifying and studying such signatures is an overarching goal in quantum thermodynamics \cite{Uzdin15,Vinjanampathy16}. We first characterize the genuinely quantum non-equilibrium steady-state (NESS) of our driven-dissipative system. Focusing on the First Law, we then evidence new quantum components in the energetic flows, and combine these findings to study the behavior of these energy flows in the quantum NESS. Finally, we analyze the impact of the coarse-graining time on the emergence of such signatures.

\subsection{Quantum Non-equilibrium Dynamics} \label{s:QNESS}

\begin{figure}
\begin{center}
\includegraphics[width=0.48\textwidth]{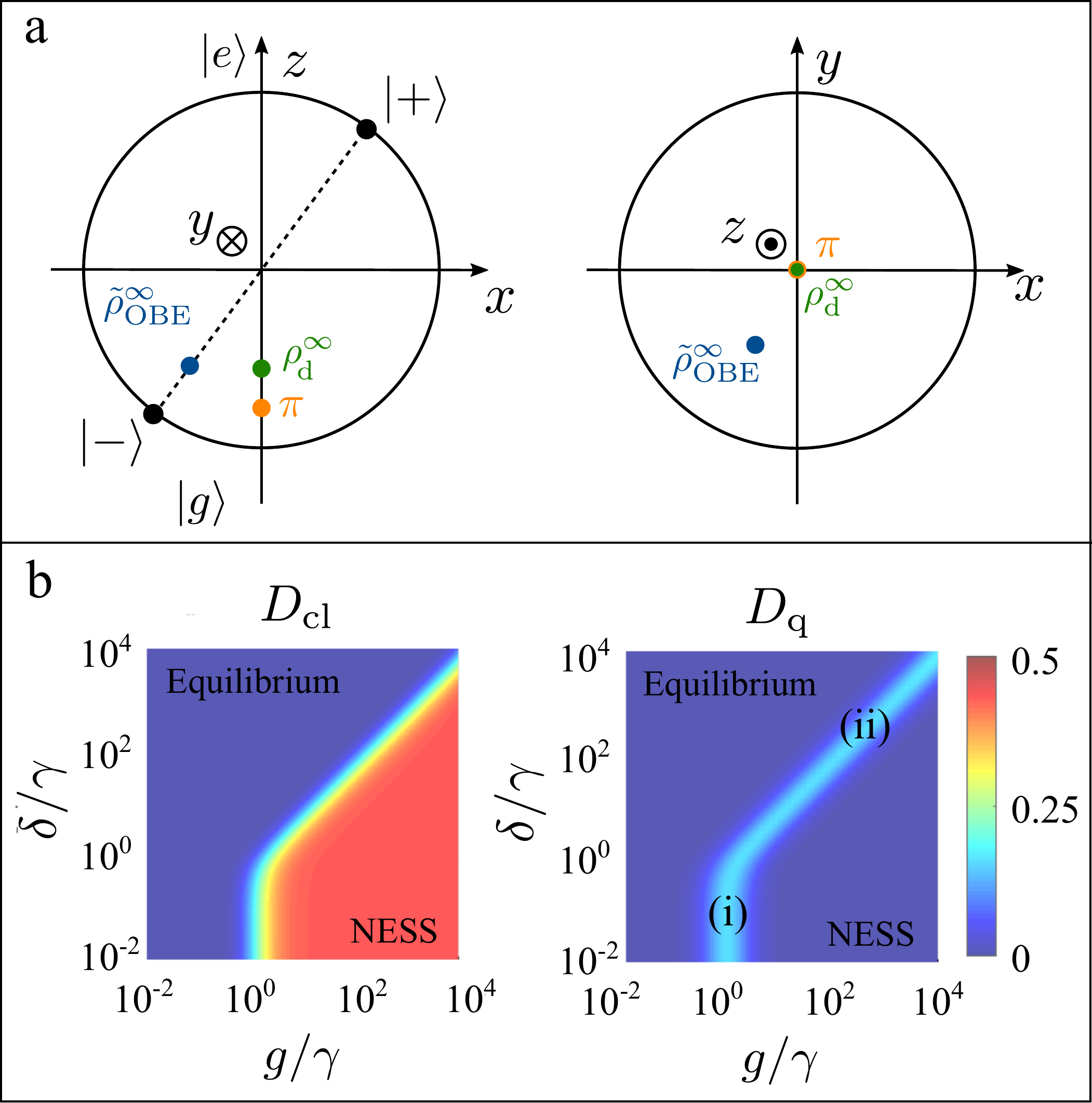}
\end{center}
\caption{\textbf{a}: Bloch sphere representations in the rotating frame. Blue dot: Steady-state $\tilde{\rho}^\infty_\text{OBE}$, orange dot: thermal equilibrium state $\pi$, green dot: diagonal projection $\rho_\text{d}^\infty$ (See text). Left: (x0z) plane; Right: (x0y) plane. Parameters: $\delta = g = \gamma$, $\gamma/\omega_\text{L} = \numprint{1e-5}$, $\hbar\omega_\text{L}/k_\text{B}T = 2$. \textbf{b}: Left (Right) Relative entropy $D_\text{cl}$ ($D_\text{q}$) as a function of $(g/\gamma, \delta/\gamma)$ (See text). Parameters: $\gamma/\omega_\text{L} = \numprint{1e-5}$, $\hbar\omega_\text{L}/k_\text{B}T = 2$.
} 
\label{f:BlochSphere}
\end{figure}

In the classical realm, a non-equilibrium dynamics is typically produced by coupling some thermodynamic system of interest to thermal baths of different temperatures, or by driving the system in a non-adiabatic manner. When transposed in the quantum regime, the latter situation can generate genuinely quantum non-equilibrium situations, the resonance fluorescence modeled by the OBE providing a canonical example. To make this point obvious, we consider the respective evolutions of the average qubit population $P_1(t) = \bra{1}\tilde\rho(t)\ket{1}= \bra{1}\rho(t)\ket{1}$ and coherence $\tilde s(t) = \bra{1}\tilde\rho(t)\ket{0} = e^{i\omega_\text{L} t}\bra{1}\rho(t)\ket{0}$. They verify:
\begin{subequations}
\bb
\dot P_1(t) &=& -\gamma((2\bar{n}+1) P_1(t)-\bar{n}) + \dfrac{ig}{2}(\tilde s(t)-\tilde s(t)^*)\nonumber\\\label{eq:OBEPe}\\
\dot {\tilde s} &=& -\left(i\delta + \dfrac{\gamma(2\bar{n}+1)}{2}\right)\tilde s(t)  + ig\left(P_1(t)-\dfrac{1}{2}\right).\nonumber\\\label{eq:OBEs}
\ee\label{eq:OBEPes}
\end{subequations}
Note that from now on, unless explicitly written otherwise, the populations and the coherences will systematically be defined in the bare qubit energy basis, which also corresponds to the eigenbasis of the thermal equilibrium state $\pi = e^{-H_\text{qb}/k_\text{B}T}/Z$. Equations \eqref{eq:OBEPe} and \eqref{eq:OBEs} capture the continuous build up of coherences, that are continuously erased by the bath. The competition between these two mechanisms gives rise to a quantum non-equilibrium steady-state (NESS) further denoted $\tilde{\rho}_\text{OBE}^\infty$ (see Fig.~\ref{f:BlochSphere}\textbf{a}). By definition, $\tilde{\rho}_\text{OBE}^\infty$ differs from $\pi$, leading to a strictly positive steady-state entropy production rate $\dot \sigma^\infty$. As a first quantum signature, $\tilde{\rho}_\text{OBE}^\infty$ carries coherences (see Appendix C for the detailed expression). This property appears to be generically true for non-adiabatically driven quantum open systems \cite{Dann18}.

To further characterize the classical and quantum features of the NESS, we have derived its relative entropy to the thermal equilibrium state $D(\tilde{\rho}_\text{OBE}^\infty||\pi)$. It splits into two components \cite{Santos19}, giving rise to two non-equilibrium features:
\bb
D(\tilde{\rho}_\text{OBE}^\infty||\pi)& =& D_\text{q}  + D_\text{cl} \\
D_\text{q} &= &D(\tilde{\rho}_\text{OBE}^\infty||\rho_\text{d}^\infty) \label{eq:Dq} \\
D_\text{cl}& =& D(\rho_\text{d}^\infty||\pi) \label{eq:Dd}.
\ee
We have introduced $\rho_\text{d}^\infty = \sum_{i=e,g} \ket{i}\bra{i}\tilde\rho^\infty_\text{OBE}\ket{i}\bra{i}$ the projection of the steady-state in the eigenbasis of the thermal equilibrium state. A strictly positive value of $D_\text{cl}$ (resp. $D_\text{q}$) signals that the steady-state populations (resp. coherences) differ from the thermal ones.  $D_\text{cl}$ and $D_\text{q}$ are plotted in Fig.~\ref{f:BlochSphere}\textbf{b} as a function of the driving strength $g$ and detuning $\delta$ for a fixed temperature $T= \hbar \omega_\text{L} /2 k_\text{B}$. As it appears on the Figure, the thermal equilibrium characterized by $D_\text{cl} = D_\text{q} = 0$ is reached as soon as the driving strength is negligible with respect to the damping rate $g \leq \gamma$ or the detuning $g \leq \delta$. Conversely, the opposite regime of NESS systematically gives rise to a steady-state population of the excited level strictly larger than the equilibrium one, and therefore $D_\text{cl}>0$.

Originally, the boundary between the NESS and the equilibrium is characterized by the presence of steady-state coherences, highlighted by $D_\text{q} > 0$. Two regimes can be distinguished. The dissipative resonant regime (i) corresponds to $\gamma \sim g \gg \vert\delta\vert$. In this case, the steady-state coherences are purely imaginary (See Appendix C), which means that the qubit's dipole and the driving field oscillate in phase quadrature. Conversely, the quasi-unitary detuned regime (ii) is defined by $g \gg \vert\delta\vert $. In this limit, the steady-state coherences are purely real, and the qubit's dipole oscillates in phase with the drive. As it appears on Fig.\ref{f:BlochSphere}\textbf{b}, the steady-state coherences are fragile and vanish as soon as the driving strength becomes too weak or too strong. In particular, the latter case is characterized by an increasing uncertainty on the phase of the Rabi oscillation, blurring out the dipole phase and the associated steady-state coherences.

\subsection{Quantum energy components}

We now search for specific quantum signatures in the First Law. To do so, we shall systematically write the thermodynamic expressions as a function of the population and coherences of the qubit's quantum state and explore the physical meaning of the obtained components.

Firstly, we rewrite the qubit internal energy $U(t) = \text{Tr}\{\rho(t)H(t)\}$ as $U(t) = U_\text{cl}(t)+U_\text{q}(t)$, where 
\bb
U_\text{cl}(t)&=&\text{Tr}\{\rho(t)H_\text{qb}\} = \hbar \omega_\text{qb} (P_1(t) - 1/2),\\
U_\text{q}(t)&=&\text{Tr}\{\rho(t)V(t)\}= \hbar g \text{Re} \tilde{s}(t),
\ee
are state functions. The splitting $H(t) = H_\text{qb}+V(t)$ was defined in Section \ref{s:Redfield}. $U_\text{cl}(t)$ is the component of the qubit's energy that is stored in the populations. Conversely, $U_\text{q}(t)$ is proportional to the real part of the coherences. In the steady-state, $U^\infty_\text{q}$ thus solely takes non negligible values at the boundary (ii) of the NESS, where it can be shown to have the opposite sign of the detuning $\delta$ (See Appendix C). $U^\infty_\text{q}$ quantifies an effective interaction between the field and the qubit, whose repulsive or attractive nature is controlled by the sign of $\delta$. This interaction has a classical interpretation in the limit of large $\vert\delta\vert/g$ where the population of the excited state is negligible. In this case, $U^\infty_\text{q}$ corresponds to the dipolar potential and is routinely exploited to trap atoms. However, the classical interpretation breaks down in the regime of quantum NESS where $g\sim \vert\delta\vert$ and the excited state population cannot be neglected. In Appendix B.5, we provide a physical interpretation of $U_\text{q}(t)$ based on the quantization of the drive. We show that $U_\text{q}(t)$ signals the presence of quantum correlations between the qubit and the drive and has no classical equivalent. This analysis confirms the role of $U_\text{q}(t)$ as an energetic quantum signature of resonance fluorescence.\\

We now study the temporal evolution of $U_\text{cl}(t)$ and $U_\text{q}(t)$. They read
\begin{equation}
d_t U_\text{cl}(t) = \dot{Q}_\text{cl} + \dot E_\text{R} \label{eq:dUcl}
\end{equation}
and
\begin{equation}
d_t U_\text{q}(t) = \dot{Q}_\text{q} + \dot{W} -  \dot E_\text{R}  . \label{eq:dUq}
\end{equation}
We have defined the classical $\dot{Q}_\text{cl}$ (resp. the quantum $\dot{Q}_\text{q}$) component of the heat flow, that respectively involve the qubit state's populations and coherences, namely
\bb
\dot Q_\text{cl}(t) &=& \text{Tr}\{{\cal L}_\text{OBE}[\rho] H_\text{qb}\}\nonumber\\
&=& - \gamma \hbar\omega_\text{qb} (\bar n + 1/2)\left(2P_1(t)-1 +\frac{1}{2\bar n+1}\right)\quad \label{eq:Qcl}
\ee
and 
\bb
\dot Q_\text{q}(t) &=& \text{Tr}\{{\cal L}_\text{OBE}[\rho] V(t)\}\label{eq:QqV}\\
&=& - \gamma \hbar g \frac{2\bar n +1}{2} \text{Re}\tilde s(t). \label{eq:Qq}
\ee
The term $\dot E_\text{R}= -(i/\hbar)\text{Tr}\{ [H, \rho] H_\text{qb}\}$ represents the reversible energy exchange between the populations and coherences induced by the driving. It is the energetic counterpart of the Rabi oscillation. The First Law is recovered by summing up Eq.~\eqref{eq:dUcl} and \eqref{eq:dUq}.

These equations reveal the structure of energy flows experienced by a driven qubit coupled to a thermal bath. From Eq.~\eqref{eq:dUq}, it appears that the work flow provided by the drive and the quantum component of the heat flow feed the energy stored in the qubit's coherences, that is quantified by $U_\text{q}(t)$. From there, the coherences are converted into populations through the exchange term flow (Eq.~\eqref{eq:dUcl}). Conversely, the energy stored in the populations $U_\text{cl}(t)$ can be converted back into coherences, or get dissipated through the classical heat flow.

The analysis above identifies $\dot{Q}_\text{q}$ as the change of $d_t U_\text{q}$ induced by the bath. Moreover, it appears from the expressions of $U_\text{q}(t)$ and $\dot{Q}_\text{q}$ that $U_\text{q}(t)\times\dot{Q}_\text{q} \leq 0$, such that this variation is always associated with a decrease of $\vert U_\text{q}\vert$, and therefore of the magnitude of the coherences. This allows us to interpret the quantum component of the heat flow as the power provided by the bath to erase the coherences. This physical interpretation is reminiscent of the concept of quantum heat introduced by some of us \cite{Elouard17Role}, that quantifies the energy flow associated to measurement back-action. Here, the action the bath on the qubit can be understood as a generalized measurement in the bare qubit basis. The energy change of the qubit associated with this measurement naturally emerges from our thermodynamic framework as it contributes to the total heat flow provided by the bath.

 \subsection{Steady-state quantum signatures} 
 
 \begin{figure}
\begin{center}
\includegraphics[width=0.48\textwidth]{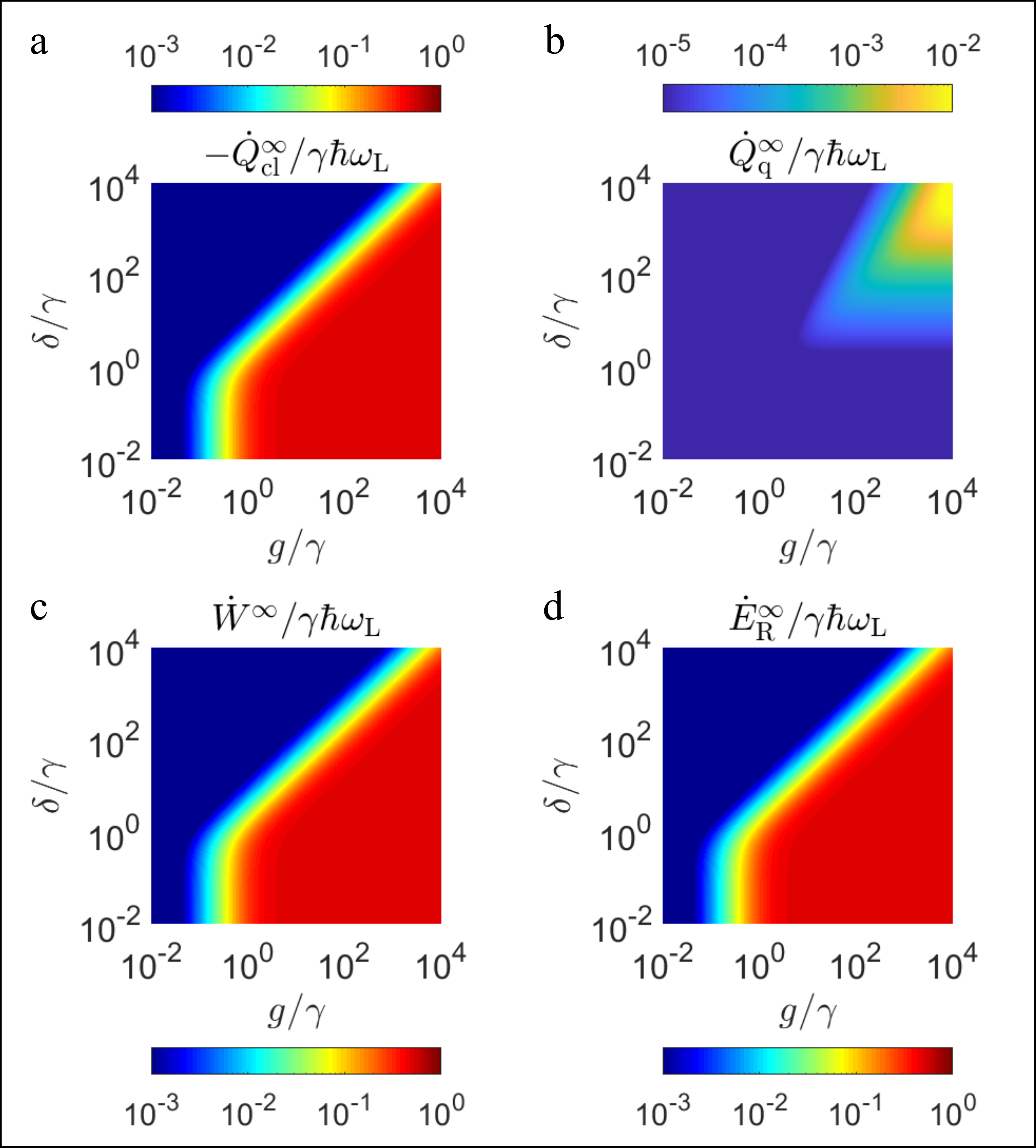}
\end{center}
\caption{Steady-state thermodynamic variables as a function of $(g/\gamma, \delta/\gamma)$. \textbf{a}, \textbf{b}: Steady-state heat flows $\dot Q^\infty_\text{cl}$ and $\dot Q^\infty_\text{q}$. \textbf{c}: Steady-state work flow $\dot W^\infty$. \textbf{d}: Steady-state coherent energy exchange rate $\dot E_\text{R}^\infty$ (See texte). Parameters: $\gamma/\omega_\text{L} = \numprint{1e-5}$, $\hbar\omega_\text{L}/k_\text{B}T = 2$. 
\label{f:QNESS}
}
\end{figure}

$ U _\text{cl}$ and $U_\text{q}$ being state--functions, their evolution equations Eq. \eqref{eq:dUcl} and \eqref{eq:dUq} become in the steady-state 

\begin{equation}
0 = \dot{Q}^\infty_\text{cl} + \dot E_R^\infty \label{eq:dUclss}
\end{equation}
and
\begin{equation}
0 = \dot{Q}^\infty_\text{q} + \dot{W}^\infty -  \dot E_R^\infty \label{eq:dUqss}
\end{equation}
The steady-state values of the various flows are denoted with $^\infty$. The First Law reads eventually
\begin{equation}
\dot W^\infty = - \dot Q_\text{cl}^\infty - \dot Q_\text{q}^\infty, \label{eq:1stLawss}
\end{equation}
highlighting the steady-state conversion of the work flow into heat. The rate of entropy production in the NESS is quantified by $\dot \sigma^\infty = -\dot Q^\infty / T = \dot{W}^\infty /T$ which is always positive. $\dot Q^\infty_\text{cl}$, $\dot Q^\infty_\text{q}$, $\dot W^\infty$ and $\dot E_R^\infty$ are respectively plotted in Fig.~\ref{f:QNESS}\textbf{a}, \textbf{b}, \textbf{c} and \textbf{d}. They all vanish at thermal equilibrium. 

It is enlightening to consider the orders of magnitude of the energetic steady-state contributions in the non-equilibrium regimes of parameters identified above. We first consider the resonant regime of NESS defined as $g \gtrsim \gamma,\vert\delta\vert$. In this case, the quantum component of the heat flow vanishes, such that the work flow is entirely converted into the classical heat flow. Both quantities scale like $\gamma \hbar \omega_L$, which characterizes the rate of photons of the drive that are absorbed by the qubit and and emitted into the bath. This conversion is mediated by the coherent exchange term $E^\infty_R=\hbar g \omega_\text{L} \text{Im}(\tilde{s}^\infty)$. When $g\gg\gamma$, the large value of $g$ compensates for the vanishingly small value of the coherences $\text{Im} \tilde s^\infty \sim \gamma/g$ such that this term remains finite, verifying $\dot E^\infty_R = \dot W^\infty = - \dot Q_\text{cl}^\infty$.  This analysis reveals that even if the quantum coherences take negligible values, the non-equilibrium dynamics entirely relies on a coherent mechanism of energy exchange. As such, the NESS it gives rise to has no classical interpretation.

Conversely, the detuned regime of the NESS is characterized by $ \vert \delta \vert \gg g$. Here the quantum component of the heat flow $\dot Q^\infty_\text{q}$ takes non-zero values, revealing the steady-state power provided by the bath to erase the coherences. We stress that this cost can be positive ($\delta>0$) or negative ($\delta <0$), such that $\dot Q^\infty_\text{q}$ respectively decreases or increases the net total heat flow. In the same regime, the entropy production rate $\dot{\sigma}^\infty$ therefore contains a genuinely quantum contribution $\dot{\sigma}_\text{q}^\infty = -\dot{Q}_\text{q}^\infty/T$ associated with coherence erasure, which has the sign of $-\delta$, i.e. increasing or decreasing the irreversible nature of the process. Note that as mentioned above, $\dot{\sigma}^\infty$ always remains strictly positive. 

Finally, we stress that since $\dot Q^\infty_\text{q}$ scales like $\hbar g$, its value remains largely dominated by the classical heat flow $\dot Q^\infty_\text{cl}$. Therefore, unveiling quantum thermodynamic signatures like the quantum heat flow is expected to be experimentally challenging. In this spirit, one may wonder whether it is even possible to evidence $\dot Q^\infty_\text{q}$. In Ref.~\cite{QuantumSwitch}, a feasible experiment is designed to single out this quantum component by isolating it from the classical contribution $\dot Q_\text{cl}$. The proposal involves a driven qubit coupled to two thermal baths of different temperatures. The parameters are chosen such that the steady-state qubit populations match the thermal equilibrium value associated with one of the two baths, keeping a finite value of the steady-state coherences. 
Such regime holds the promise to experimentally evidence the quantum heat as the total heat flow exchanged with that bath. \\

\subsection{Impact of the coarse-graining time}

The thermodynamic analysis of the OBE conducted above revealed quantum signatures in the resonance fluorescence, that had been overlooked so far: The coherent nature of the work-to-heat conversion in any regime, and the existence of a quantum component in the heat flow. These findings were allowed by the short coarse-graining $\Delta t_\text{OBE}$, which resolves the temporal evolution associated of Rabi oscillations. Conversely, many Rabi oscillations and many relaxation events occur during the long coarse-graining time $\Delta t_\text{FME}$ used in the derivation of the FME, such that coherent and incoherent processes become temporally indistinguishable. 

This averaging effect explicitly appears in the derivation of the FME from the OBE in Section \ref{s:FMEvsOBE}. The imaginary part of the steady state coherences (see Fig.~\ref{f:Comp}\textbf{c} and \textbf{f}) vanishes in the FME steady-state. The coherent energy exchange mechanism, quantified by the term $E_\text{R}^\infty \propto \text{Im}\tilde s^\infty$ is thus replaced by an effective, incoherent transfer of work provided by the drive directly into the heat flow. In the same way, the quantum component of the heat flow cannot be distinguished from the classical contribution, both being contained in $\dot Q_\text{FME}$ (see Fig.~\ref{f:CompThermo}).
Eventually, the only quantum signature in the behavior of the FME appears in the eigenbasis of the Floquet steady-state, i.e. the dressed states $\ket{\pm}$ of the qubit. These states carry coherences in the bare qubit energy eigenbasis, a property that lies at the basis of reservoir engineering techniques \cite{Lewenstein87,Murch12}. On the other hand, once expressed in the dressed states basis, the mechanism of resonance fluorescence resembles a classical non-equilibrium process by which a classical two-level system is coupled to two thermal baths of different temperatures.

\section{Conclusion and perspectives}

We have provided a so far missing framework to analyze Optical Bloch Equations in the coherent driving regime from a thermodynamic standpoint. We derived such framework from first principles, building on a microscopic description of the bath and a partial secular approximation applied consistently to the dynamical and the thermodynamic equations. When using the coarse graining time leading to the OBE, the newly derived thermodynamic variables are shown to explicitly depend on the quantum coherence in the qubit's energy basis, providing so far overlooked quantum thermodynamic signatures. Our framework opens the way to a full thermodynamic analysis of the deep coherent regime of light-matter interaction, which has remained unexplored so far. Our results can be directly tested in setups where thermal noise is the main source of dephasing, e.g. involving qubits based on superconducting circuits \cite{Nakamura99,You07}, trapped ions \cite{Monroe95,Bruzewicz19}, or electronic states of atoms \cite{Bloch08,Saffman09}.\\

Let us conclude with some important remarks.  
Firstly, the method presented here can straightforwardly be extended to more complex cases, including multilevel driven systems, or multipartite interacting systems in the strong (but not ultra-strong) coupling regime. These situations correspond to a system Hamiltonian of the form $H= H_0 + V$, where all transition frequencies $\omega_k$ in $H_0$ largely overcome the matrix elements $v_{ij}$ of $V$ which can include small detunings, weak couplings or periodic drivings. Conversely, the matrix elements of $V$ can have the same magnitude as the bath-induced transitions between the non-degenerate eigenstates of $H_0$. In the spirit of the findings presented above, one can choose a coarse-graining time $\Delta t$ fulfilling $\omega_k,\tau_\text{c}^{-1} \gg \Delta t^{-1} \gg v_{ij}$, for all ${i,j,k}$, where $\tau_\text{c}$ stands for the correlation time of the bath. This leads to a local master equation, namely, its non-unitary part does not involve the matrix elements of $V$. This method provides an analytical support to the numerical studies of the accuracy of local master equations \cite{McCauley20}. Moreover just like the OBE, the obtained master equation has the potential to reveal genuinely quantum NESS. The strategy presented in Section \ref{s:Thermo} can indeed be applied to prove that for this generic class of master equations, the work and heat flows have the same expressions as in the case of an adiabatic drive. Deriving and studying the thermodynamics of such local master equations constitutes a natural follow-up of the present paper.

Secondly, our theory provides a framework to study the energetic cost of quantum computation. As a matter of fact, it accurately describes the thermodynamics of a single qubit gate in the presence of thermal noise, allowing us to assess the minimal work needed to activate it or the entropy produced during these non-unitary operation. This paves the way towards the study of complex architectures such as Noisy Intermediate-Scale Quantum (NISQ) computers \cite{Arute19}, or large scale, fault tolerant quantum computers involving more gates and other types of quantum noises.

Finally, our framework provides a new playground to study quantum signatures in quantum engines, where work extraction is based on stimulated emission. In this novel context, coherent driving has been shown to give rise to quantum advantages such as power boosts \cite{Uzdin15,Klatzow19}. Such studies have focused so far on adiabatic or weak periodic drivings. Our approach unlocks the possibility to investigate such engines in the regime of strong periodic driving at short timescales where the effect of coherences and consequently their advantages is expected to be important.\\

\section*{Acknowledgements}
We thank Gregory Bulnes Cuetara and Benjamin Huard for enlightening discussions. C.E. was supported by the US Department of Energy (DOE), Office of Sciences, Basic Energy Sciences (BES), under Award No. DE-SC0017890. Work by A.A. and M.E was supported by grant number FQXi-IAF19-05 from the Foundational Questions Institute Fund, a donor advised fund of Silicon Valley Community Foundation. M.E. is also funded by the European Research Council (project NanoThermo, ERC-2015-CoG Agreement No.~681456). This work was conducted in part at the KITP, a facility supported by the U.S. National Science Foundation under Grant No. NSF PHY-1748958. 

\section*{Appendix A: Lamb and light shifts}

The master equation Eq.~\eqref{eq:Redfield}, once $H_{\cal SB}^I(t)$ was replaced with its explicit expression contains term involving the quantity
\bb
K(\nu) = \frac{1}{\hbar^2}\int_0^\infty  d\tau e^{i\nu\tau}\moy{R^I(\tau)R^I(0)}\label{d:Knu},
\ee
whose real part $\text{Re}K(\nu) \equiv G(\nu)/2$ is the spectral density defined in Eq.~\eqref{d:Gnu}. This identity uses that the correlation function is invariant under time translation:

\bb
\moy{R^I(t+\tau)R^I(t)} &=& \sum_k g_k^2 \Big[e^{-i\omega_k\tau}N(\omega_k)\nonumber\\
&&\quad\quad\quad +  e^{i\omega_k\tau}(N(\omega_k) + 1)\Big]\nonumber\\
&=& \moy{R^I(\tau)R^I(0)}.
\ee

The imaginary part
\bb
D(\nu) &=& 2\text{Im}K(\nu)\nonumber\\
&=& 2{\cal P} \sum_k g_k^2\Big[\frac{N(\omega_k)}{\nu-\omega_k} + \frac{N(\omega_k) + 1}{\nu+\omega_k}\Big],
\ee
where ${\cal P}$ stands for the Cauchy principal value, contributes to $\dot \rho$ with a term:
\bb
&&\dot\rho_\text{sh.}  = \frac{i}{4}\sum_{\omega} D(-\omega+\omega_\text{L})\Big(\sigma_-(\omega)\rho\sigma_+-\sigma_+\sigma_-(\omega)\rho - \text{H.c.}\Big)\nonumber\\
&&+ \frac{i}{4}\sum_{\omega} D(-\omega-\omega_\text{L})\Big(\sigma_+(\omega)\rho\sigma_--\sigma_-\sigma_+(\omega)\rho - \text{H.c.}\Big).\quad\quad
\ee

Using that $\Omega \ll \omega_\text{L}$ and neglecting the dependence of the system-bath coupling and thermal occupation function on $\omega$ as in Section \ref{s:DerivationOBE}, we obtain:
\bb
&&\dot\rho_\text{sh.} \simeq -\frac{i}{4} D(\omega_\text{L})\Big[\sigma_+\sigma_-,\rho\Big]- \frac{i}{4}D(-\omega_\text{L})\Big[\sigma_-\sigma_+,\rho\Big].\quad
\ee

This contribution can be interpreted as a renormalization of the qubit transition frequency by an amount:
\bb
\delta_\text{sh.} = \frac{D(\omega_\text{L})-D(-\omega_\text{L})}{2}.
\ee

This value accounts both for the Lamb shift (value for $\bar n =0$) and the dynamic Stark shift, or light shift (temperature dependent part). Values of these shifts are computed in Ref.~\cite{Farley81} for typical Rydberg qubits, finding typical values in the kHz range on top of GHz transitions.

\section*{Appendix B: Interpretation if the FME in the radiative cascade} \label{rad-casc}

\subsection*{B.1 Quantum model of the drive}

Here we go beyond the classical approximation. We model the drive as a coherent field $\ket{\alpha}$ injected in a mode of frequency $\omega_L$, that contains a large number of photons $|\alpha|^2 \gg 1$ \cite{CCT}. In the following we shall take $\alpha$ real without loss of generality. Using the same notations as in the main text, the Hamiltonian of the joint qubit-field system reads 
\bb
H_\text{tot}= \frac{\hbar \omega_\text{qb}}{2} \sigma_z + \hbar g_0 (a^\dagger \sigma + \sigma^\dagger a) + \hbar \omega_L a^\dagger a,
\ee
where $a$ is the mode annihilation operator (note that the Rotating Wave Approximation was applied as in the main text \cite{CCT}). The semi-classical description used in the main text is recovered by studying the dynamics in the interaction picture with respect to the field's free Hamiltonian. We then neglect the action of the qubit on the field and trace over the field's mode to derive the equation evolution of the reduced qubit dynamics. It is ruled by the time-dependent Hamiltonian $H(t)$ with $g = g_0  \alpha$. 

\subsection*{B.2 Dynamics and effective transition frequencies}

\begin{figure}[t]
 \begin{center}
 \includegraphics[width=0.48\textwidth]{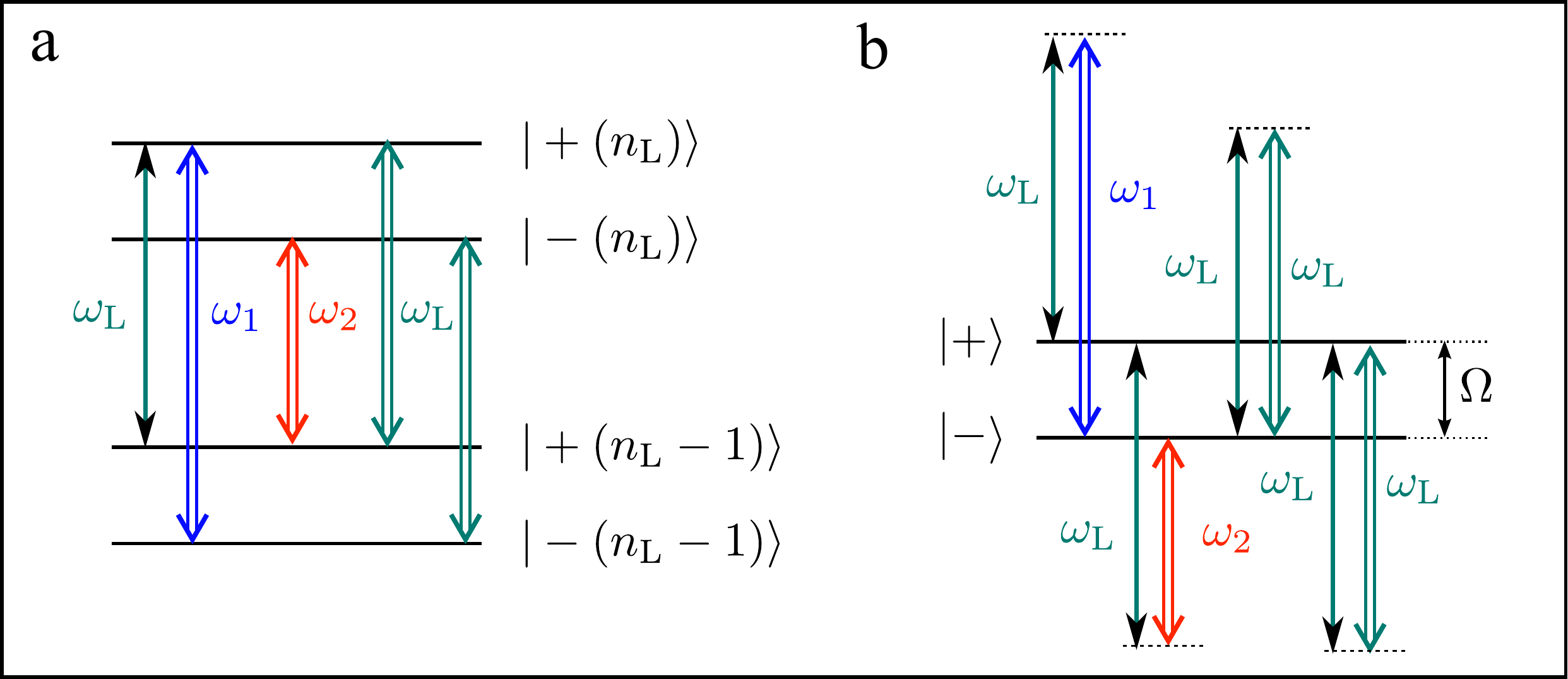}
 \end{center}
 \caption[Interpretation of Floquet dynamics]{Interpretation of Floquet dynamics. \textbf{a}: Energy diagram of the joint system field-qubit when the field is modeled by a cavity mode (radiative cascade picture). The frequencies $\omega_{1,2}$ and $\omega_\text{L}$ appear as the allowed transitions between these joint eigenstates. \textbf{b}: The FME involve transitions between the dressed states that are assisted by the driving field which can provide or take a single quantum of energy $\hbar\omega_\text{L}$. This description can be deduced from the radiative cascade model by tracing over the field subspace.\label{f:RadiativeCasc}}
 \end{figure}

An intuitive explanation for the origin of the three frequencies involved in the FME can be obtained from this quantum model. The cavity-qubit energy diagram associated with Hamiltonian $H_\text{tot}$ is a quasi-periodic ladder involving two-state manifolds $\{\ket{+(n_\text{L})},\ket{-(n_\text{L})}\}$, labeled by the number of photons in the cavity $n_\text{L} \sim \vert\alpha\vert^2 \gg 1$. In this classical limit of the field, the bath induces transitions between the levels of manifolds $n_\text{L}$ and $n_\text{L}\pm1$ which are separated by energy splittings $\hbar\omega_{1,2}$ and $\hbar\omega_\text{L}$. The reduced qubit dynamics thus consists in transitions induced by the bath between the dressed states $\{\ket{+},\ket{-}\}$ introduced in Eqs.~\eqref{d:plus}-\eqref{d:minus} and linked to the cavity-qubit energy eigenstates via $\ket{\pm}\bra{\pm} = \sum_{n_\text{L}} \ket{\pm(n_\text{L})}\bra{\pm(n_\text{L})}$. As these transitions link subspaces of different values of $n_\text{L}$, the driving field provides or takes a photon of frequency $\omega_\text{L}$ each time a photon is emitted or absorbed. The frequencies $\omega_{1,2}$ and $\omega_\text{L}$ are then formed as the the sum of the transition frequency in the driving field $\pm\omega_\text{L}$ and that in the qubit $\omega\in\{0,\pm\Omega\}$ (see Fig.~\ref{f:RadiativeCasc}). This picture also explains the apparent negative temperature of channel ${\cal L}_2$ (resulting in $\gamma_{2,\downarrow}\leq \gamma_{2,\uparrow}$) as an artifact of the reduction of the periodic ladder onto the two-level qubit subspace. This channel corresponds to transitions between states $\ket{-(n_\text{L})}$ and $\ket{+(n_\text{L}-1)}$ such that the excitation of the qubit $\ket{-}\to\ket{+}$ is associated with the \emph{emission} of a photon in the bath while the laser provides an excitation. We finally emphasize that while in the main text we started our study with a classical description of the driving field (only via driving Hamiltonian $V(t)$), the role of the field in assisting the transition between dressed states is still captured and can be tracked back to be a consequence of the time-oscillating coefficients in $V(t)$.

\subsection*{B.3 Apparition of the non-conservative force}

The partial trace over the state of the field allowing to retrieve the FME from the radiative cascade reduces a pseudo-periodic problem -- the infinite ladder of the field-qubit energy diagram (see Fig.~\ref{f:RadiativeCasc}\textbf{b}) -- onto an effective periodic description in a smaller configuration space -- the two dressed states of the qubit (see Fig.~\ref{f:RadiativeCasc}\textbf{a}). Such reduction procedure is known in classical thermodynamics to introduce non-conservative forces \cite{Esposito15,Raz16,Fogedby17,Rao18}, associated with an energy flow that must be accounted as work despite its apparent dissipative nature. Here, this energy flow simply accounts for the variation of the field photon number associated with each transition. 

\subsection*{B.4 Constant terms in the heat flow and entropy production}

The constant term proportional to $\hbar\omega_\text{L}$ in the heat flow (last line of Eq.~\eqref{eq:QFME}) can be seen as another witness of the radiative cascade \cite{CCT}. Indeed, it comes from the so-called  ``pseudo-transitions'' \cite{Langemeyer14}, i.e. the transitions $\ket{+(n_\text{L})}\to \ket{+(n_\text{L}\pm 1)}$ and $\ket{-(n_\text{L})}\to\ket{-(n_\text{L}\pm 1)}$, which in the reduced two-level description only results in a pure dephasing rate with no apparent change of populations, but as it is clear in the cavity-qubit space are associated with a transfer of energy from the driving field directly into the bath. Similarly, they lead to the last term in Eq.~\eqref{eq:sigmaFME}.

\subsection*{B.5 Two-level qubit and drive as a closed system}

We now use this model to gain insight in the energy contribution $U_\text{q}$. Starting from the product state $\ket{0,\alpha}$ and in the absence of coupling to the bath, the qubit and the field get weakly entangled \cite{Gea-Banacloche90}. Namely, their joint state at time $t$ reads 
${\ket{\Psi(t)} = \sqrt{P_0(t)}\ket{0, \psi_0(t)}+\sqrt{P_1(t)}\ket{1,\psi_1(t)} } $ where $\ket{\psi_1(t)}$ (resp. $\ket{\psi_0(t)}$) is the normalized field state correlated with the qubit excited (resp. ground) state, verifying in the classical limit $\ket{\psi_{0/1}(t)} \sim \ket{\alpha(t)} = \ket{\alpha e^{-i\omega_\text{L}t}}$. We have introduced the respective population of the excited and ground state $P_1(t)$ and $P_0(t)$. Conversely, the mean number of photons in the cavity reads $\bra{\Psi(t)} a^\dagger a \ket{\Psi(t)} = P_1(t) n_1(t) + P_0(t) n_0(t)$, where $n_i$ stands for the cavity number of photons if the qubit is in the state $i = 0,1$.

We now consider the operator $\hat{N} = \sigma^\dagger \sigma + a^\dagger a$. It verifies $[\hat{N}, H_\text{tot}] = 0$, such that the number of excitations is a conserved quantity along the joint dynamics. It yields $P_1(t)(n_e(t)+1) + P_0(t)n_g(t) = |\alpha|^2$. Conversely, the total energy for the joint system reads $U_\text{tot}(t) = U_\text{cl}(t)+ U_\text{f}(t) + V(t)$. $U_\text{cl}(t)$ and $U_\text{f}(t)$ are local energy terms. Just like in the main text, $U_\text{cl}(t) = \hbar \omega_\text{qb} (P_1(t)-1/2)$ while $U_\text{f}(t) = \hbar \omega_L (P_1(t)n_e(t)+ P_0(t) n_g(t))$. The correlation term reads $V(t) = \hbar \omega_\text{qb} \sqrt{P_0(t)P_1(t)} \text{Re}[\bra{\psi_g(t)}a^\dagger \ket{\psi_1(t)}] $ where 
$\text{Re}$ stands for the real part. In the classical limit, $\text{Re}[\bra{\psi_0(t)}a^\dagger \ket{\psi_1(t)}] \sim \alpha$ and $\text{Re}(\tilde{s}(t)) \simeq \sqrt{P_1(t)P_0(t)}$, such that it appears that $V(t)$ equals $U_\text{q}(t)$, i.e. the qubit energy component stored in the coherences. 

We can finally write the expression of energy conservation in the classical limit. It yields $U_\text{tot} /\hbar =  \omega_\text{qb} (P_1(t) -1/2) + \hbar \omega_L (P_1(t)n_1(t)+ P_0(t) n_0(t)) + U_\text{q}(t) = -\omega_\text{qb} /2 + \omega_L N(0)$. Denoting as $\delta = \omega_\text{qb} - \omega_L$ and taking into account that the number of excitations is conserved, we finally get 
\bb
\hbar \delta P_1(t) + U_\text{q}(t) = 0
\ee
$U_\text{q}(t)$ thus appears as the necessary energetic component that ensures the compatibility of the excitation number and energy conservation laws.

\section*{Appendix C: FME vs OBE, steady-state comparison} \label{FME-OBE}

\subsection*{Characterization of the steady-state}

The two master equations corresponding to the OBE and FME have very different forms and different conditions of validity. However, it exists a regime of parameters such that both choices of coarse-grainings are valid, which is:
\bb
\omega_\text{qb},\omega_\text{L}, \tau_\text{c}^{-1} \gg \Omega \gg \gamma,\gamma\bar n.\label{eq:CommonRegime}
\ee

In order to compare the predictions of both methods, we look at the steady-state values of the population of the excited state $P_1 = \bra{1}\tilde \rho^\infty\ket{1}$ and of the coherence amplitude $\tilde s= \bra{1}\tilde\rho^\infty\ket{0}$. For the OBE, we get (from Eq.\eqref{eq:OBEPes} by setting the time-derivative to $0$ and solving for $P_1$ and $\tilde s$):
\begin{subequations}
\bb
P^\infty_1 &=& \dfrac{1}{2 \bar{n}+1}\left( \bar{n} + \dfrac{1/2}{1+ 2\dfrac{\delta^2}{g^2}+ \dfrac{\gamma^2(2 \bar{n}+1)^2}{2g^2}}\right)\label{eq:Pess}\\
\tilde s^\infty &=& -\dfrac{\dfrac{\delta}{g(2n+1)} + i \dfrac{\gamma}{2g}}{1+ 2\dfrac{\delta^2}{g^2}+ \dfrac{\gamma^2(2 \bar{n}+1)^2}{2g^2}},\label{eq:sss}
\ee\label{eq:ss}
\end{subequations}
while we use the definitions of the dressed states Eqs.~\eqref{d:plus}-\eqref{d:minus} to deduce that in the case of the FME:
\bb
P_1^\infty&=& \frac{1}{2}+\frac{\delta}{2\Omega}(2\tilde P_+^\infty-1)\\
\tilde s^\infty&=&\frac{g}{2\Omega}(2\tilde P_+^\infty-1),
\ee
where $P_+^\infty$ is given in Eq.~\eqref{eq:Ppss}. 

\subsection*{Deviation between the OBE and FME steady-states in the common regime of validity}

In order to assess analytically the correspondence between the steady-state predicted by OBE and FME approach in the common regime of validity, we make an expansion of the difference between the expressions for the steady-state values of $P_1$, $\text{Re}\tilde s$ and $\text{Im}\tilde s$ for small values of $\gamma/g$ and $\hbar\Omega/k_\text{B}T$ and $\vert\delta\vert/\omega_\text{L}$. Denoting $\Delta X = X^\text{FME}-X^\text{OBE}$ for $X \equiv P_1$, $\text{Re}\tilde s$ or $\text{Im}\tilde s$, we obtain up to first order in $\frac{\gamma}{g}$, $\frac{\hbar\Omega}{k_\text{B}T}$ and $\frac{\hbar\omega_\text{L}}{k_\text{B}T}$:
\bb
\Delta P_1 &\simeq& \frac{\delta}{\omega_\text{L}} \frac{2e^{\frac{\hbar\omega_\text{L}}{k_\text{B}T}}\frac{\hbar\omega_\text{L}}{k_\text{B}T}}{\left(1+e^{\hbar\omega_\text{L}/k_\text{B}T}\right)^2}\frac{\delta^2}{g^2+2\delta^2}\\
\Delta \text{Re}\tilde s &\simeq& \frac{\delta}{\omega_\text{L}} \frac{\frac{\hbar\omega_\text{L}}{k_\text{B}T}}{1+\cosh(\frac{\hbar\omega_\text{L}}{k_\text{B}T})}\frac{g \delta}{g^2+2\delta^2} \\
\Delta \text{Im}\tilde s &\simeq& \frac{\gamma}{2g}\frac{g^2}{g^2+2\delta^2} .
\ee


Using 
\bb
\frac{\delta^2}{g^2+2\delta^2},\,\,\,\frac{\delta g}{g^2+2\delta^2},\,\,\,\frac{g^2}{g^2+2\delta^2} \leq 1
\ee
and 
\bb
\frac{2e^{\frac{\hbar\omega_\text{L}}{k_\text{B}T}}\frac{\hbar\omega_\text{L}}{k_\text{B}T}}{\left(1+e^{\hbar\omega_\text{L}/k_\text{B}T}\right)^2},\,\,\, \frac{\frac{\hbar\omega_\text{L}}{k_\text{B}T}}{1+\cosh(\frac{\hbar\omega_\text{L}}{k_\text{B}T})} \leq \frac{1}{2},
\ee
it is straightforward to conclude that in the the discrepancy between OBE and FME steady-state values is negligible in the limit where $\gamma/g$, $\hbar\Omega/k_\text{B}T$ and $\vert\delta\vert/\omega_\text{L}$ go to zero, which are three conditions required to have both FME and OBE descriptions to be valid.

\end{document}